\documentclass[final,3p,times,twocolumn]{elsarticle}

\usepackage{epsfig}

\usepackage{amssymb}

\usepackage{lineno}

\usepackage{caption}
\usepackage{subcaption}
\usepackage{bm}
\usepackage{amsmath}
\usepackage{float}
\usepackage{tabularray}
\usepackage{hyperref}
\usepackage{xcolor}

\journal{Nuclear Instruments and Methods in Physics Research A}

\makeatletter
\def\ps@pprintTitle{%
   \def\@oddhead{}%
   \def\@evenhead{}%
   \def\@oddfoot{}%
   \def\@evenfoot{}}
\makeatother

\begin{document}

\begin{frontmatter}



\title{Response Matrix Estimation in Unfolding Differential Cross Sections}


\author[cmu_stat]{Huanbiao Zhu}
\author[pisa]{Andrea Carlo Marini}
\author[cmu_stat]{Mikael Kuusela}
\author[cmu_stat,cmu_mld]{Larry Wasserman}

\affiliation[cmu_stat]{organization={Department of Statistics and Data Science, Carnegie Mellon University},%
            city={Pittsburgh}, 
            state={PA},
            country={USA}}
\affiliation[pisa]{organization={Dipartimento di Fisica, Università di Pisa},
            city={Pisa},
            country={Italy}}
\affiliation[cmu_mld]{organization={Machine Learning Department, Carnegie Mellon University},
            city={Pittsburgh}, 
            state={PA},
            country={USA}}            

\begin{abstract}
The \textit{unfolding problem} in particle physics is to make inferences about the true particle spectrum based on smeared observations from a detector. This is an ill-posed inverse problem, where small changes in the smeared distribution can lead to large fluctuations in the unfolded distribution.  
The forward operator is the response matrix
which models the detector response.
In practice, the forward operator is rarely known analytically and is instead estimated using Monte Carlo simulation. 
This raises the question of how to best estimate the response matrix and what impact this estimation has on the unfolded solutions.
In most analyses at the LHC, response matrix estimation is done by binning the true and smeared events and counting the propagation of events between the bins. However, this 
approach can result in a noisy estimate of the response matrix, especially with a small Monte Carlo sample size. Unexpectedly, we also find that the noise in the estimated response matrix can inadvertently regularize the problem. As an alternative, we propose to estimate the response matrix through the use of conditional density estimation of the response kernel in the unbinned setting followed by binning this estimator. Using simulation studies, we investigate the performance of the two approaches.
\end{abstract}




\end{frontmatter}


\section{Introduction}
\label{sec:intro}
\begin{figure}[h]
	\centering
	\begin{subfigure}{.4\textwidth}
		\centering
		\includegraphics[width=.8\linewidth]{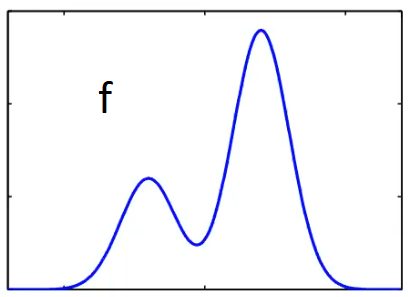}
		\caption{true distribution f}
		\label{fig:true_dist}
	\end{subfigure}
	\begin{subfigure}{.4\textwidth}
		\centering
		\includegraphics[width=.8\linewidth]{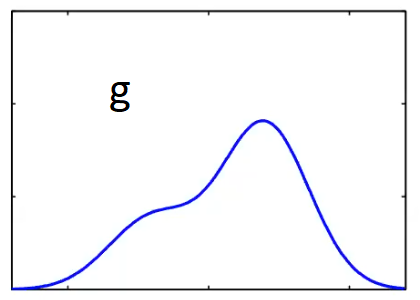}
		\caption{smeared distribution g}
		\label{fig:smeared_dist}
	\end{subfigure}
	\caption{Unfolding is the process of inferring the true distribution $f$ from the smeared distribution $g$.}
	\label{fig:unfolding}
\end{figure}

The \textit{unfolding problem} arises in the statistical analysis of data at the Large Hadron Collider (LHC) at CERN. 
In LHC experiments, it is often of interest to learn the distribution (or differential cross section) $f$ of some physical quantity measured by the detectors, e.g., the energy, mass or production angle of the particles produced in the proton-proton collisions. Due to the finite resolution of the detectors, however, the observed data follow a smeared distribution $g$, rather than $f$ itself. This detector smearing maps the true distribution $f$ to the measured distribution $g$, as illustrated in Figure~\ref{fig:unfolding}. \textit{Unfolding} refers to the process of inferring $f$ from $g$ (for details, see, e.g., \cite{Blobel2013}).

The relationship between $f$ and $g$ is given by the following integral equation:
\begin{equation}
g(y) = \int_Tk(y,x)f(x)dx, \quad y\in S,
\end{equation}
where $T$ and $S$ are the true (particle-level) space and  the smeared (detector-level) space, respectively, and $k(y,x)=p_{Y|X}(y|x)$ is the \textit{response kernel} that models the detector smearing. Due to computational and practical reasons, data in high-energy physics are frequently binned using histograms. This leads to the discretized formulation $\boldsymbol{\mu}=\boldsymbol{K} \boldsymbol{\lambda}$, where $\boldsymbol{\lambda}$ and $\boldsymbol{\mu}$ denote the particle and detector level histogram means and $\boldsymbol{K}$ is a \textit{response matrix} in which each entry represents bin-to-bin smearing probabilities. The detailed setup will be described in Section \ref{sec:discretization}. 

In the context of the discretized formulation, the goal is to make inferences on the particle-level histogram~$\bm{\lambda}$. This is an ill-posed inverse problem, where small changes in the measured distribution can lead to large changes in the unfolded distribution \cite{Blobel2013}. Existing unfolding approaches typically employ regularization techniques, such as Tikhonov regularization \cite{Tikhonov1963} or early stopping of the D'Agostini iteration \cite{DAgostini1995} (which is an EM algorithm), to address the ill-posed nature of the problem.

In practice, the response matrix $\bm{K}$ is usually not known analytically and needs to be estimated using detector simulations. This raises the question of how to best estimate the response matrix and what impact this estimation has on the unfolded solutions. The main contribution of this work is to introduce a new way to estimate the response matrix by first considering response kernel estimation in the unbinned setting. Additionally, we show that there are nontrivial, and sometimes unexpected, interactions between the response matrix estimators and the quality of the unfolded solutions. 

This paper is organized as follows: In Section~\ref{sec:unfolding_overview}, a brief overview of unfolding methodology is given in the context of the binned setting. Next, in Section~\ref{sec:matrix_estimation},
we describe several methods for estimating the response matrix. Then, in Section~\ref{sec:compare_matrix}, the response matrix estimators are compared using a simulation study of unfolding the inclusive jet transverse momentum. In Section~\ref{sec:effect_on_unfold_spectrum}, the effects of the response matrix estimators on the unfolded solution are investigated. In Section~\ref{sec:Drell-Yan}, we apply the different methods on simulated Drell-Yan events. Finally, in Section~\ref{sec:discussion}, we summarize our findings and discuss some open problems.

\section{Overview of unfolding methodology}
\label{sec:unfolding_overview}
\subsection{Discretization}
\label{sec:discretization}
In most unfolding analyses at the LHC, the data are modeled as indirectly observed Poisson point processes \cite{Kuusela2016}. However, instead of working on the continuous scale directly, the processes are binned using histograms. The formulation is as follows. Let $M$ be the Poisson point process representing the true particle-level spectrum and $N$ the Poisson point process representing the detector-level spectrum. Let $f$ and $g$ be the uniquely defined intensity functions for $M$ and $N$, respectively. Then, by definition, a particle-level value $X$ and a detector-level measurement $Y$ have probability density functions
\begin{equation}
	p_X(x) = \frac{f(x)}{\lambda(T)}, \;\; p_Y(y) = \frac{g(y)}{\mu(S)},
\end{equation}
where $\lambda(T)=\int_{T}f(x)dx$ and $\mu(S)=\int_{S}g(y)dy$. Next, we discretize the Poisson processes by partitioning the true space $T$ into bins $\{T_j\}_{j=1}^n$ and smeared space $S$ into bins $\{S_i\}_{i=1}^m$. This implies $M(T_j)\sim\operatorname{Poisson}(\lambda(T_j))$
and $N(S_i)\sim\operatorname{Poisson}(\mu(S_i))$, i.e., each bin count follows a Poisson distribution with the expected count given by the mean measure within the bin. Define the vectors $\bm{\lambda}=[\lambda(T_1),...,\lambda(T_n)]$ and $\bm{\mu}=[\mu(S_1),...,\mu(S_m)]$. Then, for any $i\in[m]$,
\begin{equation}
\begin{aligned}
	\mu(S_i) &= \int_{y\in S_i}g(y)dy 	
        = \int_{y\in S_i}\int_{x\in T}k(y,x)f(x)dxdy \\
	&= \int_{y\in S_i}\sum_{j=1}^{n}\int_{x\in T_j}k(y,x)f(x)dydx \\
	&= \sum_{j=1}^n\int_{y\in S_i}\int_{x\in T_j}k(y,x)f(x) \lambda(T_j)^{-1}\lambda(T_j)\,dx\,dy\\
	&= \sum_{j=1}^{n}K_{ij}\lambda(T_j),
\end{aligned}
\end{equation}
where $K_{ij}$ is defined as
\begin{equation}
\begin{aligned}
	K_{ij} &= \int_{y\in S_i}\int_{x\in T_j}k(y,x)f(x) dxdy \lambda(T_j)^{-1} \\
	&= \frac{\int_{y\in S_i}\int_{x\in T_j}k(y,x)f(x) dxdy}{\int_{x\in T_j} f(x)dx}.
\end{aligned}
\label{eq:kernel_to_matrix}
\end{equation}
Defining vectors $\bm{\mu} = (\mu(S_i))_{i=1}^m$ and $\bm{\lambda} = (\lambda(T_j))_{j=1}^n$, this formulates a linear system for the discretized problem,
\begin{equation}
	\label{discretized LS}
	\bm{\mu} = \bm{K}\bm{\lambda},
\end{equation}
where $\bm{K}$ is the $m \times n$ \textit{response matrix} with entries $K_{ij}$. Each entry of $\bm{K}$ represents the conditional probability $K_{ij}=P(\text{observed value in bin }i\, | \, \text{true value in bin }j)$. Then, the particle-level and detector-level histograms can be represented by two random vectors $\mathbf{x}$ and $\mathbf{y}$ such that 
\begin{equation}
    \begin{aligned}
        \mathbf{x} &\sim \text{Poisson}(\bm{\lambda}), \\
        \mathbf{y} &\sim \text{Poisson}(\bm{\mu}).
    \end{aligned}
    \label{eq:data_generating_process}
\end{equation}
The goal of unfolding is to make inferences on $\bm{\lambda}$ given the detector observation $\mathbf{y}$. In this work, we focus on point estimation of $\bm{\lambda}$ using an estimator $\hat{\bm{\lambda}}$.

\subsection{Unfolded point estimation}

In practice, we may use a normal approximation to the Poisson distributed bin counts when each bin contains a large enough number of events \cite{Kuusela2016}. This allows us to re-express the detector-level histograms in Equation (\ref{eq:data_generating_process}) as follows:
\begin{align}
\label{eq:normal_approx}
    \mathbf{y} = \bm{K}\bm{\lambda} + \bm{\epsilon}.
\end{align}
Here $\bm{\epsilon}\sim N(\bm{0},\bm{\Sigma})$ and $\bm{\Sigma}=\text{diag}(\bm{\mu})$ is the covariance matrix, with the diagonal elements equal to the expected counts within each smeared bin. Additionally, we consider 
the Cholesky decomposition of the covariance matrix $\bm{\Sigma}=\bm{LL}^{\top}$ and re-express (\ref{eq:normal_approx}) by
\begin{align}
\label{eq:cholesky}
    \bm{L}^{-1}\mathbf{y} &= \bm{L}^{-1}\bm{K}\bm{\lambda} + \bm{\eta}.
\end{align}
Since $\bm{\eta} \sim N(\bm{0}, \bm{L}^{-1}\bm{\Sigma}(\bm{L}^{-1})^{\top})$ and $\bm{L}^{-1}\bm{\Sigma}(\bm{L}^{-1})^{\top} = \bm{I}$, it follows that $\bm{\eta} \sim N(\bm{0}, \bm{I})$. Therefore, since $\bm{\mu}$ is accurately estimable using $\mathbf{y}$ and
the Cholesky decomposition always exists for the covariance matrix, we may replace $\bm{K} \hookrightarrow \bm{L}^{-1}\bm{K}$ and assume without loss of generality that $\bm{\epsilon}\sim N(\bm{0},\bm{I})$. Under this assumption and assuming that $\bm{K}$ has full column rank, the least squares estimator of $\bm{\lambda}$ is
\begin{equation}
    \label{eq:least-squares solution}
    \hat{\bm{\lambda}}_{ls} = (\bm{K}^{\top}\bm{K})^{-1}\bm{K}^\top\mathbf{y}.
\end{equation}
Statistically, the least-squares estimator is unbiased since 
\begin{equation}
\mathbb{E}[\hat{\bm{\lambda}}_{ls}]=
(\bm{K}^{\top}\bm{K})^{-1}\bm{K}^\top\bm{K}\bm{\lambda}=\bm{\lambda}.
\end{equation}
However, in practice, this estimator 
is unsatisfactory since the response matrix $\bm{K}$ is severely ill-conditioned, in which case $\hat{\bm{\lambda}}_{ls}$ contains large unphysical oscillations \cite{Blobel2013, Kuusela2016}. Therefore, some form of regularization is needed to mitigate the ill-posedness.

One of the common regularization methods used in LHC data analysis is Tikhonov regularization \cite{Tikhonov1963} and its variants \cite{Hoecker1996, Schmitt2012}, which add an $\ell_2$ penalty term to solve the penalized least squares problem
\begin{equation}
    \label{eq:Tikhonov}
    \begin{aligned}
        \hat{\bm{\lambda}}_{tik} &= \text{argmin}_{\bm{\lambda}} \left\{\|\mathbf{y}-\bm{K}\bm{\lambda}\|_2^2+\delta\|\bm{\lambda}\|^2_2\right\}\\
        &=(\bm{K}^{\top}\bm{K}+\delta\bm{I})^{-1}\bm{K}^\top\mathbf{y},
    \end{aligned}
\end{equation}
where $\delta>0$ is a regularization strength. Another common way to estimate $\bm{\lambda}$ is by D'Agostini iteration \cite{DAgostini1995}, also commonly referred to as Iterative Bayesian Unfolding (IBU). D'Agostini iteration is an EM 
(expectation-maximization)
algorithm with early stopping \cite{Kuusela2012}. Specifically, given an initial guess $\hat{\bm{\lambda}}^{(0)}>\bm{0}$, iteratively compute
\begin{equation}
    \label{eq:EM}
    \hat{\lambda}_j^{(r+1)} = \frac{\hat{\lambda}_j^{(r)}}{\sum_i{K}_{ij}}\sum_{i}\frac{{K}_{ij}{y}_i}{\sum_{l}{K}_{il}\hat{\lambda}_l^{(r)}}, \;\;\; j=1,\ldots,n,
\end{equation}
and, after $r$ iterations, the solution is given by $\hat{\bm{\lambda}}^{(r)}_{em}=(\hat{\lambda}_1^{(r)},\ldots,\hat{\lambda}_n^{(r)})$. As $r\rightarrow\infty$, it can be shown that $\hat{\bm{\lambda}}^{(r)}_{em}$ converges to the maximum likelihood estimate of $\bm{\lambda}$. Also, since each step of the iteration increases the likelihood monotonically, stopping early in the iterations regularizes the solution toward $\hat{\bm{\lambda}}^{(0)}$.

\section{Response matrix estimation}
\label{sec:matrix_estimation}
If the response matrix $\bm{K}$ is known, then we can proceed to compute the regularized estimators $\hat{\bm{\lambda}}_{tik}$ and $\hat{\bm{\lambda}}_{em}$ directly as in Equations~\eqref{eq:Tikhonov} and \eqref{eq:EM}. However, a closed form expression of the response matrix $\bm{K}$ is typically not available, but the matrix rather needs to be estimated from a dataset $\{(X_1,Y_1),.\ldots,(X_n,Y_n)\}$ produced using a Monte Carlo simulation of the physical process and the particle detector \cite{Blobel2013,Kuusela2016,Stanleyetal2022}. Here, $X_i$ represents the true physical quantities sampled from $p_{X}(x)$ and $Y_i$ represents the corresponding detector measurements sampled from $p_{Y|X}(y|X_i)$. The estimation of $\bm{K}$ introduces a source of statistical error and the first question is how to estimate the response matrix in a sensible way. (In reality, both $p_{X}$ and $p_{Y|X}$ might be misspecified in the Monte Carlo generator but we will ignore that source of error here and only focus on the error caused by the estimation of $\bm{K}$.)

\subsection{Histogram estimator}
\label{sec:naive_estimator}
Traditionally in LHC data analysis, the response matrix is estimated by binning. That is, conditioning on the true events in a given bin, count the numbers of corresponding detector measurements in each bin after smearing. Specifically, the estimator of the $(i,j)$th element of the matrix is the proportion of the Monte Carlo events originating from bin $j$ that have been captured by the detector in bin $i$, i.e.,
\begin{equation}
    \hat{K}^{hist}_{ij} = \frac{\#(X,Y) \text{'s such that $Y$ in bin $i$ and $X$ in bin $j$}}{\#X \text{'s in bin $j$}}.
\end{equation}
The histogram estimator is unbiased but can be noisy, especially when the number of events in the bin is small. For example, when unfolding a steeply falling spectrum, the relative error in $\hat{K}^{hist}_{ij}$ increases toward the tail of the spectrum where there are exponentially fewer events. This motivates us to seek a more refined method to estimate the response matrix.

\subsection{Response kernel estimation}

Instead of estimating the response matrix directly by binning the events, recall that the response matrix $\bm{K}$ depends on the response kernel $k$ through the integral in Equation \eqref{eq:kernel_to_matrix}. Therefore, a reliable estimate of $\bm{K}$ can be achieved by first obtaining an accurate estimate of the response kernel $k$. Moreover, since estimating $k$ non-parametrically requires some form of smoothing, this potentially provides a smoother estimator of $\bm{K}$ than the histogram estimator.

Hence, our first goal is to estimate the response kernel $k(y,x)$. Mathematically, $k(y,x)=p_{Y|X}(y|x)$ is just the conditional density of $Y$ given $X$. In practice, we do not know the form of $p_{Y|X}$, but only have pairs of observations $(X_1,Y_1),\ldots,(X_n,Y_n)$ from the Monte Carlo generator. Therefore, the problem is essentially that of estimating the conditional density $p_{Y|X}$ based on $\{(X_i,Y_i)\}_{i=1}^n$. In the following subsections, we discuss various non-parametric methods to perform \emph{conditional density estimation} (CDE) of $p_{Y|X}$. 
In Section \ref{sec:compare_matrix}, we show that this can lead to better estimates of the response matrix than by just binning the events directly.

\subsubsection{Kernel regression method} \label{sec:kernel cde}
Estimating the conditional density can be reduced to a non-parametric regression problem. Let $K(\cdot)$ be a symmetric non-negative density function. Denote $f(y|x):=\mathbb{E}[K_h(Y-y)|X=x]$, where $K_h(\cdot)=K(\cdot/h)/h$ and $h>0$ is the bandwidth parameter that controls the degree of smoothing (see, e.g., \cite{silverman1986density, scott2015multivariate}). Then, by Taylor expansion,
\begin{equation}
\begin{aligned}
	f(y_0|x) &= \int K_h(y'-y_0)p_{Y|X}(y'|x)dy' \\
    &= \int K(y)p_{Y|X}(y_0+hy|x)dy \\
    &= \int K(y)[p_{Y|X}(y_0|x)+hyp'_{Y|X}(y_0|x) + O(h^2)]dy \\
    &= p_{Y|X}(y_0|x)\int K(y)dy \\
    &\phantom{=}+ hp'_{Y|X}(y_0|x)\int yK(y)dy + O(h^2).
\end{aligned}
\end{equation}
Since $\int K(y)dy=1$ and $\int yK(y)dy=0$ by definition, as $h\rightarrow 0$, $f(y_0|x)\approx p_{Y|X}(y_0|x)$. In other words, we can view  $p_{Y|X}$ as a regression of $K_h(Y-y)$ on $X$. \citet{Rosenblatt1969} proposed to estimate the regression function $f(y|x)$ using the Nadaraya--Watson kernel smoother
\begin{equation}
\begin{aligned}
	\hat{f}_{h_1,h_2}(y|x) &= \sum_{i=1}^nw_i(x)K_{h_2}(y-Y_i),
\end{aligned}
\end{equation}
where
\begin{equation}
\begin{aligned}
	w_i(x) &= \frac{K_{h_1}(x-X_i)}{\sum_{j=1}^nK_{h_1}(x-X_j)}.
\end{aligned}
\end{equation}
This gives the kernel conditional density estimate $\hat{p}_{ks}(y|x)=\hat{f}_{h_1,h_2}(y|x)$. Notice that this depends on two bandwidths: $h_1$ controls the smoothing over $X$ and $h_2$ the smoothing over $Y$.

\subsubsection{Local linear method} 
\label{sec:local linear cde}
Suppose the second derivative of $f(y|x)$ w.r.t.~$x$ exists. In a small neighborhood of $x$, by Taylor expansion, we have
\begin{equation}
	f(y|z) \approx f(y|x) + f'(y|x)(z-x) = a+b(z-x).
\end{equation}
Estimating $f(y|x)$ is equivalent to estimating the coefficient $a$. \citet{Fan1992,Fan1996} proposed to find $a$ by minimizing
\begin{equation}
\label{eq:local_linear_optimization}
	\sum_{i=1}^n\left(K_{h_2}(y-Y_i)-a-b(X_i-x)\right)^2K_{h_1}\left(x-X_i\right)
\end{equation}
with respect to $a$ and $b$.
That is, we are estimating the least squares coefficient of the weighted local linear regression of $K_{h_2}(y-Y)$ on $X$. A direct calculation shows that the minimizer is
\begin{equation}
	\hat{f}_{h_1,h_2}(y|x)=\hat{a} = \frac{\sum_{i=1}^nw_i(x)K_{h_2}(y-Y_i)}{\sum_{i=1}^nw_i(x)}
\end{equation}
with
\begin{equation}
\begin{aligned}
	w_i(x) &= K_{h_1}\left(x-X_i\right)\left[\sum_{j=1}^n(x-X_j)^2K_{h_1}\left(x-X_j\right)- \right. \\
 &\left. (x-X_i)\sum_{j=1}^n(x-X_j)K_{h_1}\left(x-X_j\right)\right].
\end{aligned}
\end{equation}
This yields the local linear estimate $\hat{p}_{ll}(y|x)=\hat{f}_{h_1,h_2}(y|x)$. Note that if we set $b=0$ in the objective function \eqref{eq:local_linear_optimization}, then the solution to the minimization problem recovers the kernel conditional density estimate $\hat{p}_{ks}(y|x)$. Compared to the kernel method, one attractive feature of the local linear method is its design adaptive property which states that asymptotically the bias of $\hat{p}_{ll}(y|x)$ does not depend on the marginal distribution of $X$ \cite{Fan1992}. This property is particularly important because the distribution of $X$ or its parametric form is not known a priori. Moreover, in practice, the marginal distribution may fall steeply over several orders of magnitude.

\subsubsection{Bandwidth selection}
\label{sec:bandwidth}
In both the kernel and local linear methods, a pair of global bandwidths $h_1$ and $h_2$ needs to be specified. We call $h_1$ and $h_2$ global since they do not depend on the value of either $x$ or $y$. 

There are many existing methods for picking $h_1$ and $h_2$. One of the simplest and computationally most efficient ways is to find the optimal bandwidths by assuming the normality of both the conditional and marginal distributions \cite{Hyndman2001,Hyndman2002}. Although it relies on the normality assumption, this can yield reasonable results even in cases where normality does not hold \citep{scott2015multivariate}. We also find this works reasonably well in our examples.

However, using simulation studies, we found that in typical unfolding scenarios, there does not exist a single pair of bandwidths $(h_1,h_2)$ that results in good conditional density estimates for all $x$. This limitation partially arises from the fact that the amount of smearing might vary with $x$. For example, consider a normal response kernel 
$k(y,x)=p_{Y|X}(y|x)=N(\mu=x,\sigma^2=x)$. 
The variance of $k$ increases as $x$ increases, so one should not expect to find $(h_1,h_2)$ that can estimate $k$ well for all $x$. Additionally, the marginal density $p_X(x)$ typically varies over several orders of magnitude and the optimal amount of smoothing will be different between the well-observed high-density regions and the sparsely observed low-density regions. In these cases, locally adaptive bandwidths, which we describe below, can be preferable to a single pair of global bandwidths.

\subsubsection{Local kernel method}
\label{sec:local kernel cde}
To incorporate adaptive bandwidths that change as a function of $x$, we introduce moving windows to locally estimate the kernel conditional density within each window. Specifically, given a point $x$ and a window size $\delta(x)>0$, we estimate $p_{Y|X}(y|x)$ by
\begin{equation}
    \hat{p}_{lk}(y|x)=\sum_{i:|x-X_i|<\delta(x)}w_i(x)K_{h_2(x)}(y-Y_i),
\end{equation}
where
\begin{align}
    w_i(x) &= \frac{K_{h_1(x)}(x-X_i)}{\sum_{j:|x-X_j|<\delta(x)}K_{h_1(x)}(x-X_j)}.
\end{align}
Within each window, the bandwidths $h_1(x)$ and $h_2(x)$ can be found using a reference rule or other data-driven methods, but based solely on the data points inside the window. This makes the bandwidths locally adaptive, allowing one to adjust for heteroscedastic errors. Common window choices, for example, could be a fixed-size window with $\delta(x)=\delta'$ for some $\delta'>0$, or an exponential window with $\delta(x)=d_1\exp(d_2 x)$ for some $d_1 > 0$, $d_2 \in \mathbb{R}$. The exponential window is particularly useful when dealing with a steeply falling spectrum. In such cases, there are exponentially fewer data points in the upper tail of the spectrum, so exponentially growing window sizes can help account for the decreasing number of data points.

Also, it is important to distinguish between local linear and local kernel methods, despite both being termed "local". In local linear method, "local" refers to locally fitting a line, but its bandwidths $h_1,h_2$ are global that control the smoothing over $X$ and $Y$, respectively. On the other hand, in local kernel method, a separate kernel regression is fitted within each moving window, and "local" here refers to locally adaptive bandwidths $h_1(x), h_2(x)$ that vary with $x$.

\subsubsection{Location-scale model} \label{sec:location-scale} 
Aside from the kernel and local linear methods, we introduce another method based on making an additional assumption on the data-generating process. Specifically, we assume the smeared observations are generated from the following model:
\begin{align}
	Y = \mu(X) + \sigma(X)\epsilon, \label{eq:local_scale_assumption}
\end{align}
where the additive error $\epsilon$ follows some distribution with mean 0 and variance 1. $\sigma(x)$ allows for a heteroscedastic detector response whose variance depends on the conditioned $x$. This model implies that if we standardize the smeared measurement $Z=\frac{Y-\mu(X)}{\sigma(X)}$, then the standardized variable $Z$ has an identical distribution that does not depend on $X$. Again, we want to find an estimate for $p_{Y|X}(y|x)$, which in this case can be written as
\begin{align}
    p_{Y|X}(y|x) = \frac{1}{\sigma(x)}p_{\epsilon}\left(\frac{y-\mu(x)}{\sigma(x)}\right).
\end{align}

In this formulation, our approach involves several steps. First, we estimate the mean function $\mu(x)$ by regressing $Y$ nonparametrically on $X$, yielding an estimate $\hat{\mu}(x)$ along with residuals $\hat{r}_i = y_i - \hat{\mu}(x_i)$. Then, we estimate the variance function $\sigma^2(x)$ by regressing the squared residuals $\hat{r}_i^2$ on $X$, resulting in an estimate $\hat{\sigma}^2(x)$. After that, we estimate the distribution $p_{\epsilon}$ using kernel density estimation on the fitted standardized residuals $\frac{\hat{r}_i}{\hat{\sigma}(x_i)}$, obtaining an estimate $\hat{p}_{\epsilon}$.
Finally, we obtain the conditional density estimate
\begin{align}
	\hat{p}_{ls}(y|x) &= \frac{1}{\hat{\sigma}(x)}\hat{p}_{\epsilon}\left(\frac{y-\hat{\mu}(x)}{\hat{\sigma}(x)}\right).
\end{align}

A key benefit of this approach is that it accounts for the different amounts of smearing by directly estimating the variance function $\sigma^2(x)$. This avoids having to
find local bandwidths as in the local kernel method. By making a stronger modeling assumption, this method is able to make more effective use of the MC sample, but will naturally suffer from a bias if the true data-generating process differs significantly from that of Eq.~\eqref{eq:local_scale_assumption}.

\subsection{Plug-in estimator}
\label{sec:plug_in_estimator}

So far, we have considered several non-parametric CDE methods to estimate the response kernel $k$ on the unbinned space. But since our goal is to perform unfolding with binned histograms, as is most commonly done in practice, we still need to use the estimated response kernel to obtain an estimate of the response matrix $\bm{K}$.

To estimate the response matrix, a natural approach is to use a plug-in estimator by replacing the assumed known response kernel $k$ in Equation \eqref{eq:kernel_to_matrix} with the estimate $\hat{k}$. Moreover, since $\bm{K}$ depends on the true particle-level function $f$, which is unknown in the first place, a Monte Carlo ansatz $f^{MC}$, which can be misspecified from the true function $f$, is in practice used in the computation of $\bm{K}$ \cite{Stanleyetal2022}. One can estimate $f^{MC}$ using the Monte Carlo samples $X_i$. Then, the estimated response matrix has entries given by 
\begin{align}
\label{eq:estimate_K}
	\hat{K}_{ij} &= \frac{\int_{y\in S_i}\int_{x\in T_j}\hat{k}(y,x)f^{MC}(x) dxdy}{\int_{x\in T_j} f^{MC}(x)dx}.
\end{align}

$\hat{\bm{K}}$ is exposed to bias from the misspecified Monte Carlo ansatz, with increasing dependence on $f^{MC}$ as the particle-level bin size gets larger. This is referred to as wide-bin bias and one way to remedy this bias is by using finer particle-level bins to unfold and then aggregate the results into the desired wider bins \citep{Stanleyetal2022}. Since the focus of this paper is not on wide-bin bias, we will assume that $f^{MC}$ is correctly specified, i.e., $f^{MC}=f$, so we can solely focus on the implications of replacing $k$ by $\hat{k}$.

\subsection{Implementation details}
The kernel and local linear methods are implemented in the R package \href{https://cran.r-project.org/web/packages/hdrcde/index.html}{\texttt{hdrcde}} by \citet{hdrcde}. We use a Gaussian kernel and the bandwidth is selected based on the normal reference rule. The local kernel method utilizes the same package but with additional local moving windows. In our two case studies, we adopt exponentially growing window sizes to account for data sparsity in the high $p_\perp$ region. Specifically, we use windows of size $\exp(x/400)$ in the first case study (Sections~\ref{sec:compare_matrix} and \ref{sec:effect_on_unfold_spectrum}) and $\exp(x/100)$ in the second case study (Section~\ref{sec:Drell-Yan}). The window sizes differ slightly to reflect the different kinematic ranges of the two studies: 400–1000 GeV for the first case study and 20–400 GeV for the second. For the location-scale model, the mean function $\mu$ and variance function $\sigma^2$ are estimated by smoothing splines using the core R function \href{https://www.rdocumentation.org/packages/stats/versions/3.6.2/topics/smooth.spline}{\texttt{smooth.spline}}. The error distribution $\epsilon$ is estimated by kernel density estimation using the R function \href{https://www.rdocumentation.org/packages/ks/versions/1.10.7/topics/kde}{\texttt{kde}} from the \texttt{ks} package \cite{KS}. The code for all implemented methods is available at \url{https://github.com/richardzhs/response_matrix_estimation}.

\section{Comparison of response matrix estimators}
\label{sec:compare_matrix}
We will first compare the performance of response matrix estimation using the different methods from Section \ref{sec:matrix_estimation} by simulating unfolding of inclusive jets.

\begin{figure*}[t]
    \centering
    \includegraphics[width=\textwidth]{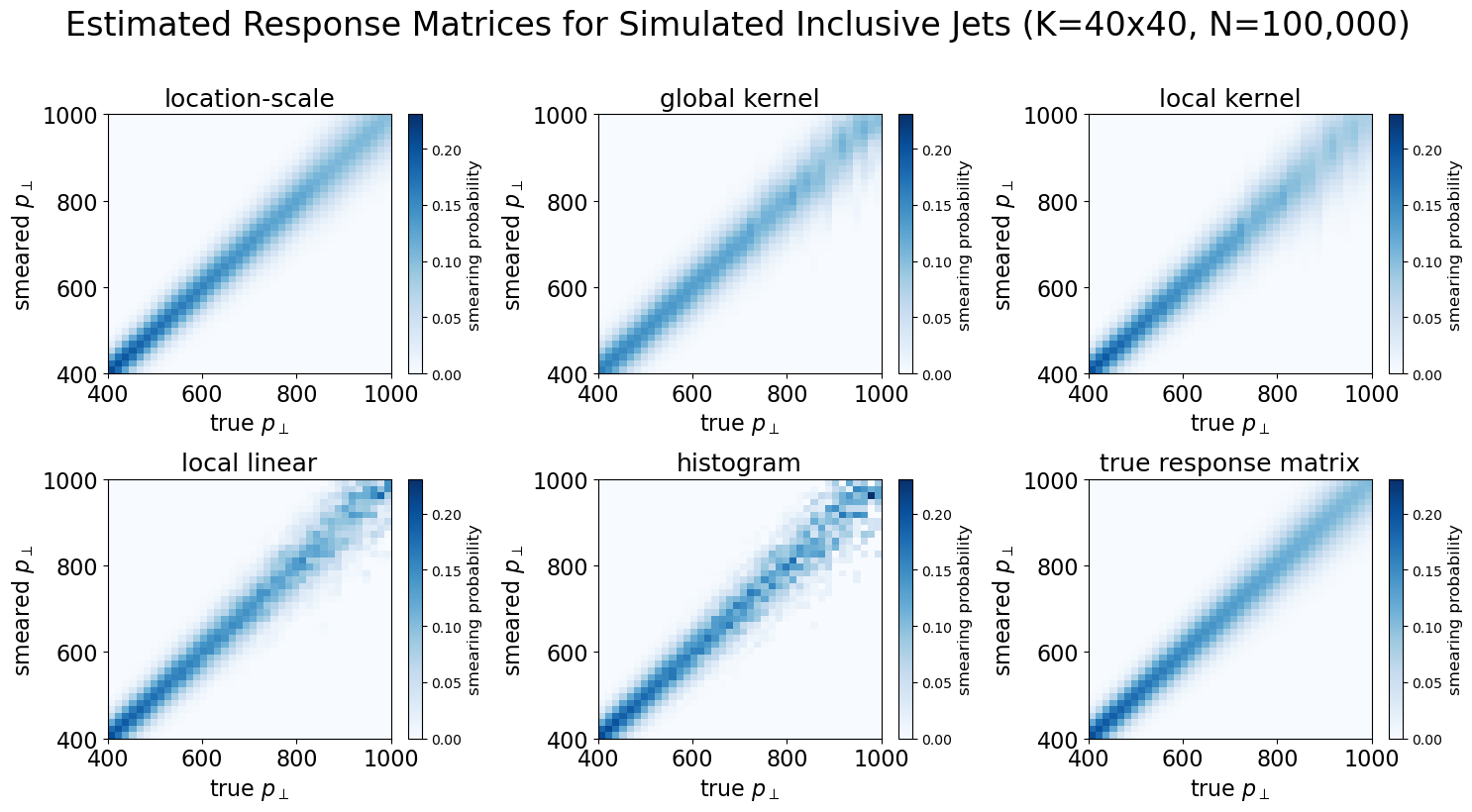}
    \caption{Estimated $40 \times 40$ response matrices using the different methods with one Monte Carlo simulation. The sample size of the Monte Carlo simulation is 100{,}000. The bottom-right heatmap is the true response matrix. The other heatmaps are the estimated response matrices with the different methods.}
    \label{fig:mat_K=40x40_n=100000}
\end{figure*}

\subsection{Simulation setup}
We simulate data to mimic unfolding of the inclusive jet transverse momentum ($p_\perp$) spectrum, which is a canonical example of a steeply falling particle spectrum. Our simulation setting is from \cite{KuuselaStark2017}, where the particle-level data are simulated using the following intensity function
\begin{align}
f\left(p_{\perp}\right)=L N_{0}\left(\frac{p_{\perp}}{\mathrm{GeV}}\right)^{-\alpha}\left(1-\frac{2}{\sqrt{s}} p_{\perp}\right)^{\beta} e^{-\gamma / p_{\perp}},
\end{align}
for $0<p_{\perp} \leq \sqrt{s}/2$. Here, $L,N_0,\alpha,\beta,\gamma,\sqrt{s}$ are parameters whose values are motivated by the setting at the LHC. The detector response is modeled as additive Gaussian noise so that $k(p_\perp',p_\perp)=N(p_\perp' \mid p_\perp,\sigma(p_\perp)^2)$ with the heteroscedastic variance satisfying
\begin{align}
\left(\frac{\sigma\left(p_{\perp}\right)}{p_{\perp}}\right)^{2}=\left(\frac{C_{1}}{\sqrt{p_{\perp}}}\right)^{2}+\left(\frac{C_{2}}{p_{\perp}}\right)^{2}+C_{3}^{2},
\end{align}
where $C_1, C_2, C_3$ are parameters depending on the detector. At the center-of-mass energy $\sqrt{s}=7\operatorname{TeV}$, realistic values for the parameters of $f$ are $L=5.1 \operatorname{fb}^{-1},\, N_0=10^{17}\operatorname{fb/GeV},\,\alpha=5,\,\beta=10$ and $\gamma=10\operatorname{GeV}$ and the parameters of $\sigma$ are set to be $C_1=1\operatorname{GeV}^{1/2}, C_2=1\operatorname{GeV},$ and $C_3=0.05$. Given these, the smeared intensity function is given by
\begin{align}
g\left(p_{\perp}^{\prime}\right)=\int_{T} N\left(p_{\perp}^{\prime} \mid p_{\perp}, \sigma\left(p_{\perp}\right)^{2}\right) f\left(p_{\perp}\right) \mathrm{d} p_{\perp}, \quad p_{\perp}^{\prime} \in S.
\end{align}
In this simulation, the particle-level and detector-level spaces are restricted to $T=S=[400 \text{ GeV},1000 \text{ GeV}]$. We partition both spaces using $m=n=40$ bins which corresponds to a response matrix of size $40\times 40$. The 7 TeV case study has been used in several previous methodological studies \citep{KuuselaStark2017, Kuusela2016}, allowing for a direct comparison with those works. However, to reflect more current experimental conditions, we also consider a 13 TeV case study in Section \ref{sec:Drell-Yan}.

\subsection{Results}
Figure~\ref{fig:mat_K=40x40_n=100000} shows the estimated response matrices for the following methods: (1) histogram: direct estimation of the response matrix via events binning (Section~\ref{sec:naive_estimator}); (2) global kernel: kernel conditional density estimation with global bandwidths (Section~\ref{sec:kernel cde}); (3) local linear: local linear conditional density estimation with global bandwidths (Section~\ref{sec:local linear cde}); (4) local kernel: kernel conditional density estimation with locally adaptive bandwidths (Section~\ref{sec:local kernel cde}); (5) location-scale: estimation of mean and variance functions assuming a location-scale model (Section~\ref{sec:location-scale}); and (6) true response matrix (for reference). The sample size to estimate the response matrix is $N=100{,}000$. Pure eye inspection indicates that the histogram estimate becomes noticeably noisier toward the sparsely sampled right tail of the spectrum compared to the CDE methods.

We quantify the performance of response matrix estimation with the different methods using the bin-wise mean absolute error (MAE)
\begin{equation}
    \label{eq:mat_mae}
    \frac{1}{M}\sum_{l=1}^M \left|\hat{K}_{ij}^{l}-K_{ij}\right|, \; \text{ for all }i\in[m], j\in[n],
\end{equation}
where $\hat{K}_{ij}^{l}$ is the estimate of the $(i,j)$th bin in the $l$th simulation.
The sample size for each Monte Carlo simulation to estimate the response matrix $\hat{\bm{K}}^{l}$ is $N=100{,}000$. The MAE is averaged over $M=1{,}000$ Monte Carlo simulations.

\begin{figure*}[t]
    \centering
    \includegraphics[width=\textwidth]{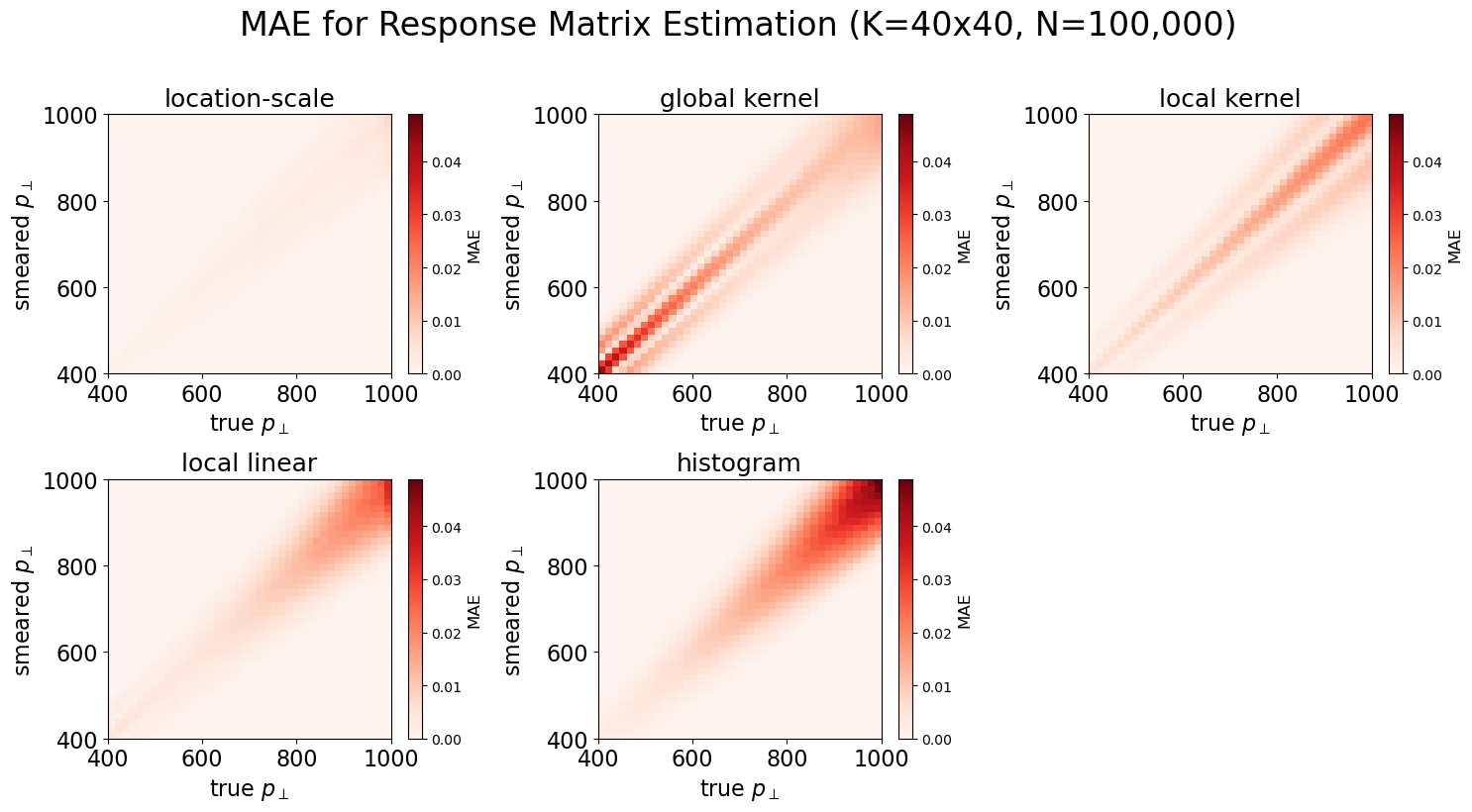}
    \caption{Bin-wise mean absolute errors for estimating the $40 \times 40$ response matrix using the different methods with 1{,}000 Monte Carlo simulations. The sample size of each Monte Carlo simulation is 100{,}000.}
    \label{fig:mat_mse_K=40x40_n=100000}
\end{figure*}

Figure \ref{fig:mat_mse_K=40x40_n=100000} shows that the histogram estimate has the largest MAE toward the upper end of the spectrum. There are also notable differences between the different CDE methods. For example, the kernel method has a larger MAE toward the lower end of the spectrum. This is due to oversmoothing in the low $p_\perp$ region. Since the response kernel has a heteroscedastic variance across $x$ and there are many fewer events in the upper tail, there does not exist a pair of global bandwidths that can provide good estimates over all conditioned $x$. On the other hand, the local kernel method with adaptive local bandwidths mitigates the large error in the lower end, although it has a slightly increased MAE in the upper end. The local linear method has a relatively large error toward the upper end, though not as large as the histogram estimate. This is due to undersmoothing in the high $p_\perp$ region. Overall, we can judge that the local linear and local kernel methods both have uniformly smaller MAEs than the histogram method, demonstrating the potential of CDE methods to produce improved estimates of the response matrix. Finally, the regression approach based on the location-scale model appears to be performing the best overall with significantly smaller MAEs, which is not surprising given that this simulation satisfies the location-scale assumption~\eqref{eq:local_scale_assumption}.

\section{Effect of the estimated response matrix on the unfolded spectrum}
\label{sec:effect_on_unfold_spectrum}
While the quality of the estimated response matrix based on the plug-in estimator (Section \ref{sec:plug_in_estimator}) is generally better compared to the naive histogram estimator (Section \ref{sec:naive_estimator}), a natural question to ask is how this affects the final unfolded spectrum. Does a better estimated response matrix automatically lead to a better unfolded point estimator? To answer this question, we compute the Tikhonov-regularized estimator \eqref{eq:Tikhonov} and D'Agostini estimator \eqref{eq:EM} of $\bm{\lambda}$ with the estimated response matrix plugged in, i.e.,
\begin{equation}
    \hat{\bm{\lambda}}_{tik} = (\hat{\bm{K}}^{\top}\hat{\bm{K}}+\delta\bm{I})^{-1}\hat{\bm{K}}^\top\mathbf{y},
\end{equation}
\begin{equation}
\begin{aligned}
    \hat{\lambda}_j^{(r+1)} &= \frac{\hat{\lambda}_j^{(r)}}{\sum_i\hat{K}_{ij}}\sum_{i}\frac{\hat{K}_{ij}y_i}{\sum_{l}\hat{K}_{il}\hat{\lambda}_l^{(r)}}, \;\;\; j=1,\ldots,n, \\
    \hat{\bm{\lambda}}^{(r)}_{em}&=(\hat{\lambda}_1^{(r)},\ldots,\hat{\lambda}_n^{(r)}).
\end{aligned}
\end{equation}

The sample size of the Monte Carlo sample for estimating the response matrix is the same as before with $N=100{,}000$. The sample size of the experimental data for constructing the histogram $\mathbf{y}$ is 1{,}000{,}000, mimicking those situations at the LHC where the available MC sample is smaller than the experimental sample. The performance of the estimators is compared using the bin-wise mean-squared error (MSE)
\begin{equation}
    \label{eq:mat_mae}
    \frac{1}{M}\sum_{l=1}^M \left(\hat{\lambda}^{l}_{j}-\lambda_j\right)^2,\; \text{  for all } j\in[n],
\end{equation}
where $\hat{\lambda}_j^{l}$ is the $j$th bin of the unfolded estimator in the $l$th simulation.
The MSE is computed using $M=1{,}000$ simulations. That is, for each simulation, we generate 100{,}000 pairs of Monte Carlo data to estimate the response matrix and 1{,}000{,}000 smeared data points to unfold.

\subsection{Numerical results for Tikhonov regularization}

\begin{figure}[t]
    \centering
    \begin{subfigure}{\linewidth}
        \centering
        \includegraphics[width=\linewidth]{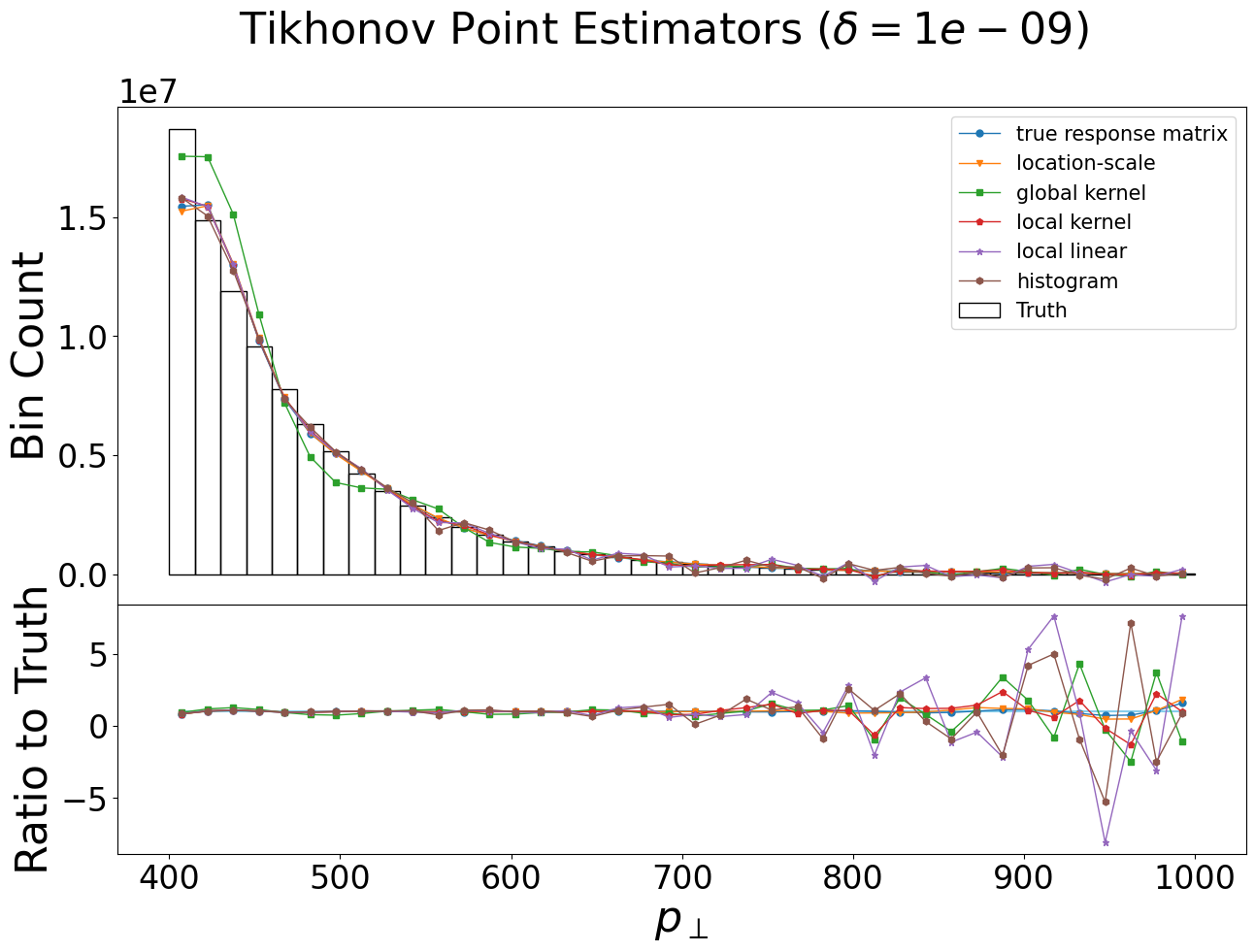}
        \caption{Tikhonov solution with $\delta=10^{-9}$.}
        \label{fig:tikhonov_sol_alpha=1e-9}
    \end{subfigure}

    \vspace{0.8em}

    \begin{subfigure}{\linewidth}
        \centering
        \includegraphics[width=\linewidth]{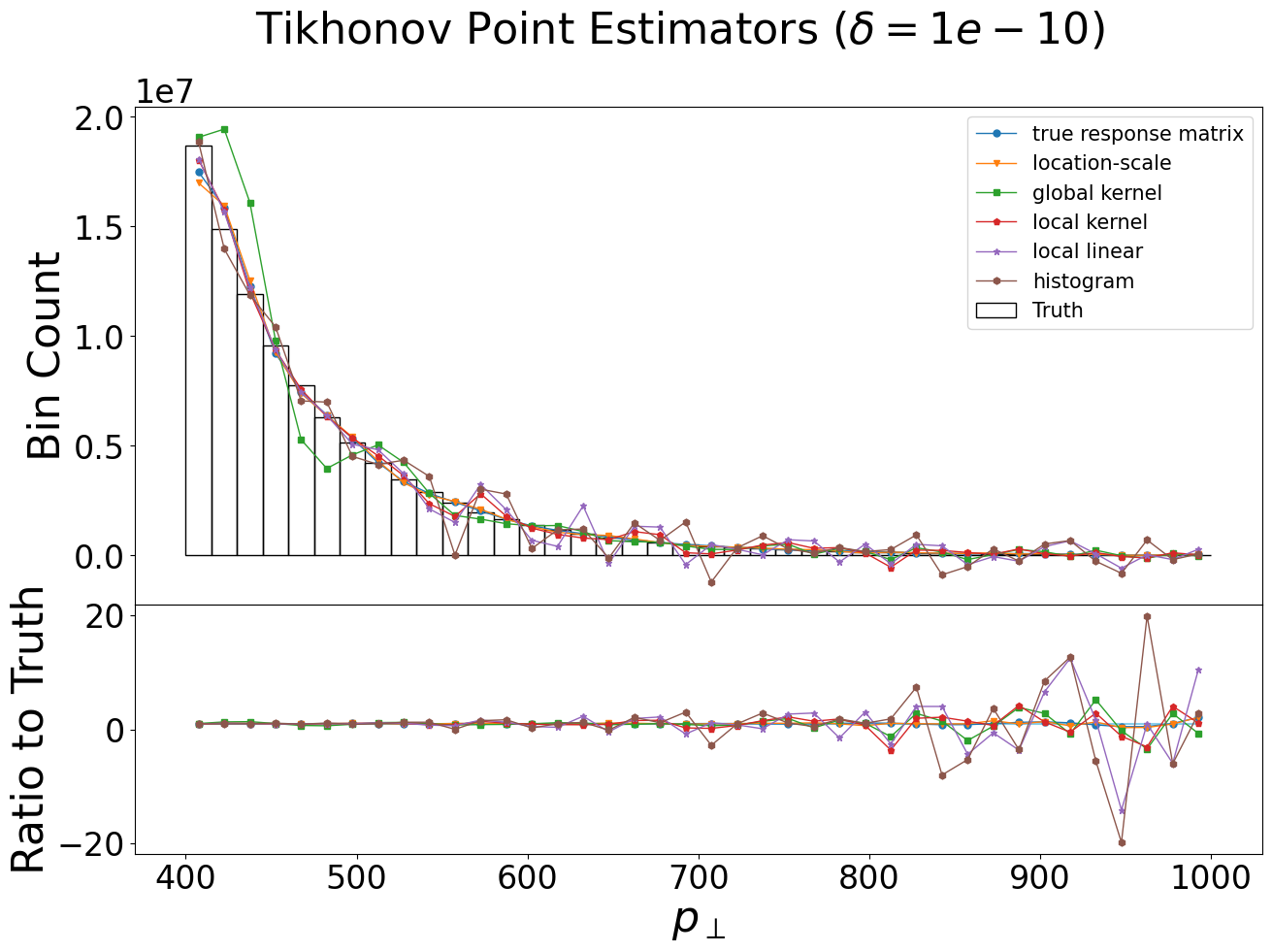}
        \caption{Tikhonov solution with $\delta=10^{-10}$.}
        \label{fig:tikhonov_sol_alpha=1e-10}
    \end{subfigure}

    \caption{
    Tikhonov solutions with regularization parameters (a) $\delta=10^{-9}$ and (b) $\delta=10^{-10}$.
    }
    \label{fig:tikhonov_solutions_1e-9_1e-10}
\end{figure}

\begin{figure}[H]
    \centering
    \includegraphics[width=\linewidth]{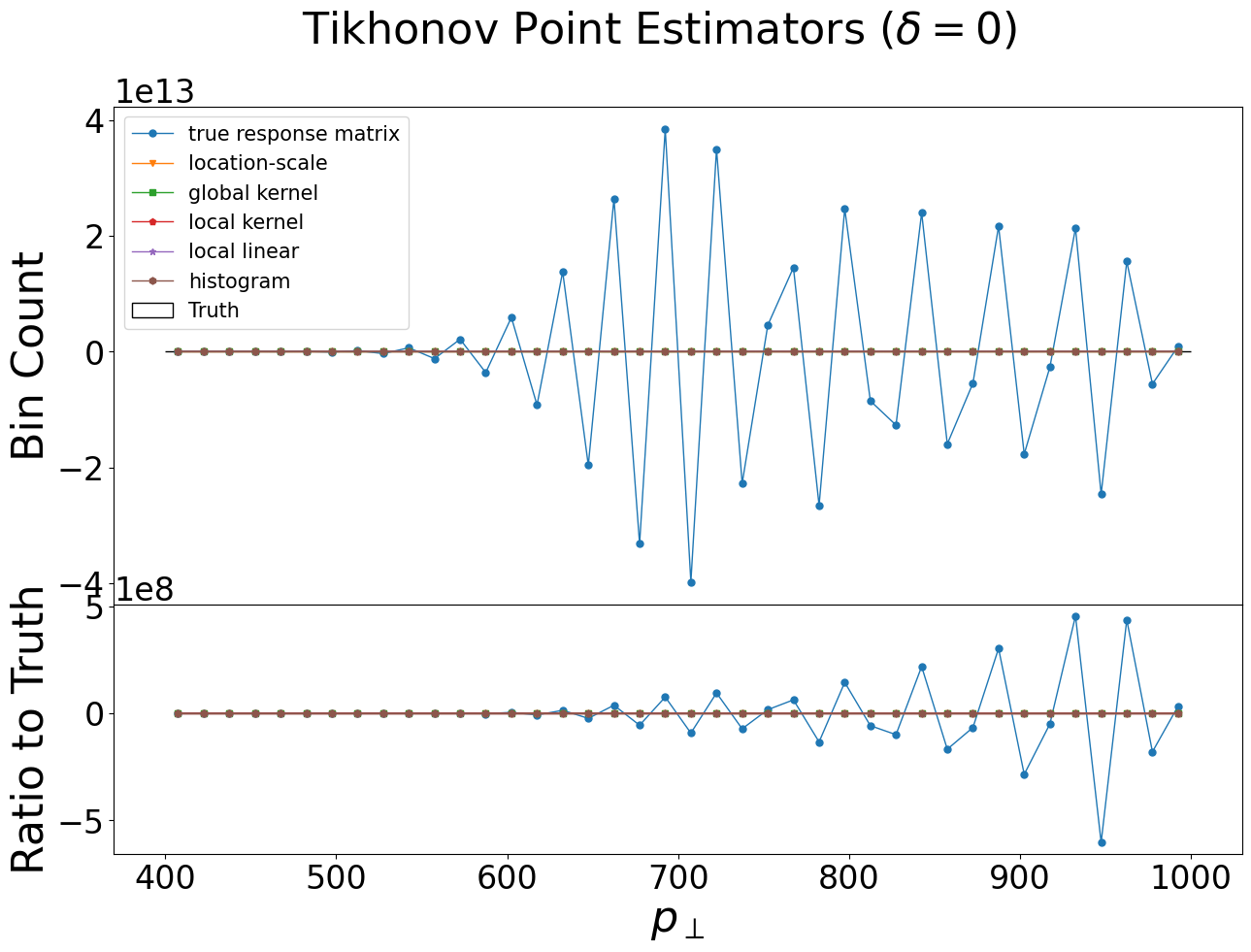}
    \caption{Unregularized least-squares solution ($\delta=0$).}
    \label{fig:tikhonov_sol_alpha=0}
\end{figure}

Figures \ref{fig:tikhonov_solutions_1e-9_1e-10}--\ref{fig:tikhonov_sol_alpha=0} illustrate one realization of unfolded solutions with Tikhonov regularization using various regularization strengths $\delta$. Figure \ref{fig:mse_tikhonov} shows the corresponding (log-scale) mean squared error (MSE), squared bias and variance for $M=1{,}000$ realizations. Rather than selecting $\delta$ according to an optimality criterion such as MSE or goodness-of-fit statistic, we intentionally consider a broad range of values, including $\delta=0$ (i.e. no regularization), in order to assess how the unfolded solution behaves across different regularization regimes. In Figures \ref{fig:mse_tikhonov_delta=1e-9} and \ref{fig:mse_tikhonov_delta=1e-10}, when $\delta=10^{-9}$ or $\delta = 10^{-10}$, the solution using the true response matrix generally has the lowest MSE, followed by the location-scale method. In contrast, the histogram estimate exhibits the highest MSE. Notice that for the high $p_\perp$ region ($p_\perp\geq 600 \text{ GeV}$), the MSE is largely driven by the variance (recall that $\text{MSE} = \text{bias}^2 + \text{variance}$). In this regime, the histogram method shows the largest variance, whereas the solution based on the true response matrix has the smallest variance, with the location–scale method also exhibiting relatively low variance compared to the other approaches. This behavior indicates that the variance of the Tikhonov-regularized solution is closely related to the quality of the estimated response matrix. Overall, we find that the relative performance of the unfolded point estimators aligns well with the relative performance of each response matrix estimation method in Section~\ref{sec:compare_matrix}, supporting the hypothesis that a more accurately estimated response matrix leads to a better unfolded solution.

\begin{table}
\centering
\caption{The median condition numbers for the estimated response matrices over $M=1{,}000$ Monte Carlo simulations using different methods. For the true response matrix, the condition number is fixed and known.}
\begin{tblr}{
  width = 0.8\linewidth,
  colspec = {Q[319]Q[319]},
  hline{1,8} = {-}{0.08em},
}
\textbf{Estimation method} & \textbf{Median condition number} \\
True                       & $1.7 \cdot 10^{17}$                    \\
Kernel                     & $3.9 \cdot 10^{7}$
\\
Local linear               & $6.7 \cdot 10^{3}$                               
\\
Local kernel               & $1.5 \cdot 10^{8}$
\\
Location-scale             & $3.9 \cdot 10^{4}$               
\\
Naive histogram            & $2.6 \cdot 10^{3}$  
\end{tblr}

\label{table:condition_number}
\end{table}

\subsubsection{Noisy matrix acts as an implicit regularizer}
\label{sec:implicit_reg}
\begin{figure}[H]
    \centering
    \begin{subfigure}{\linewidth}
        \centering
        \includegraphics[width=\linewidth]{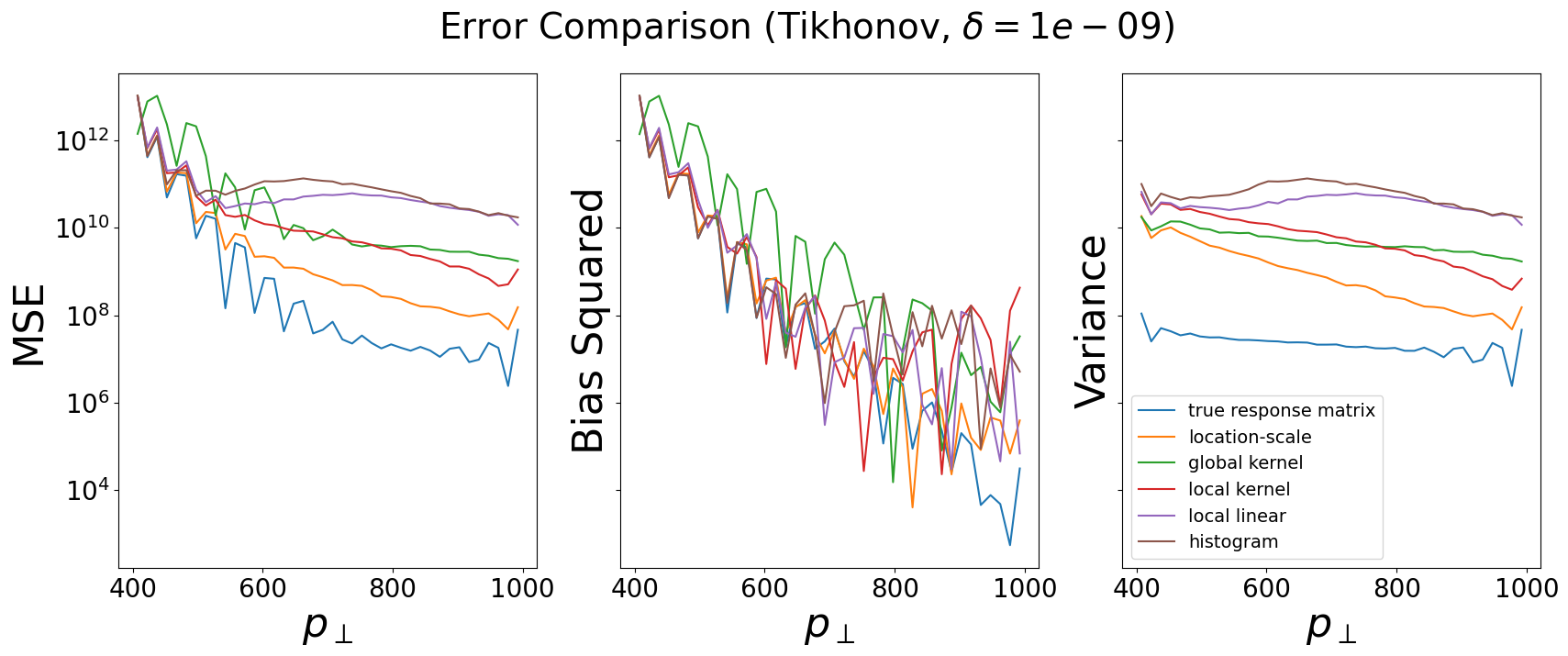}
        \caption{$\delta=10^{-9}$.}
        \label{fig:mse_tikhonov_delta=1e-9}
    \end{subfigure}

    \vspace{0.5em}

    \begin{subfigure}{\linewidth}
        \centering
        \includegraphics[width=\linewidth]{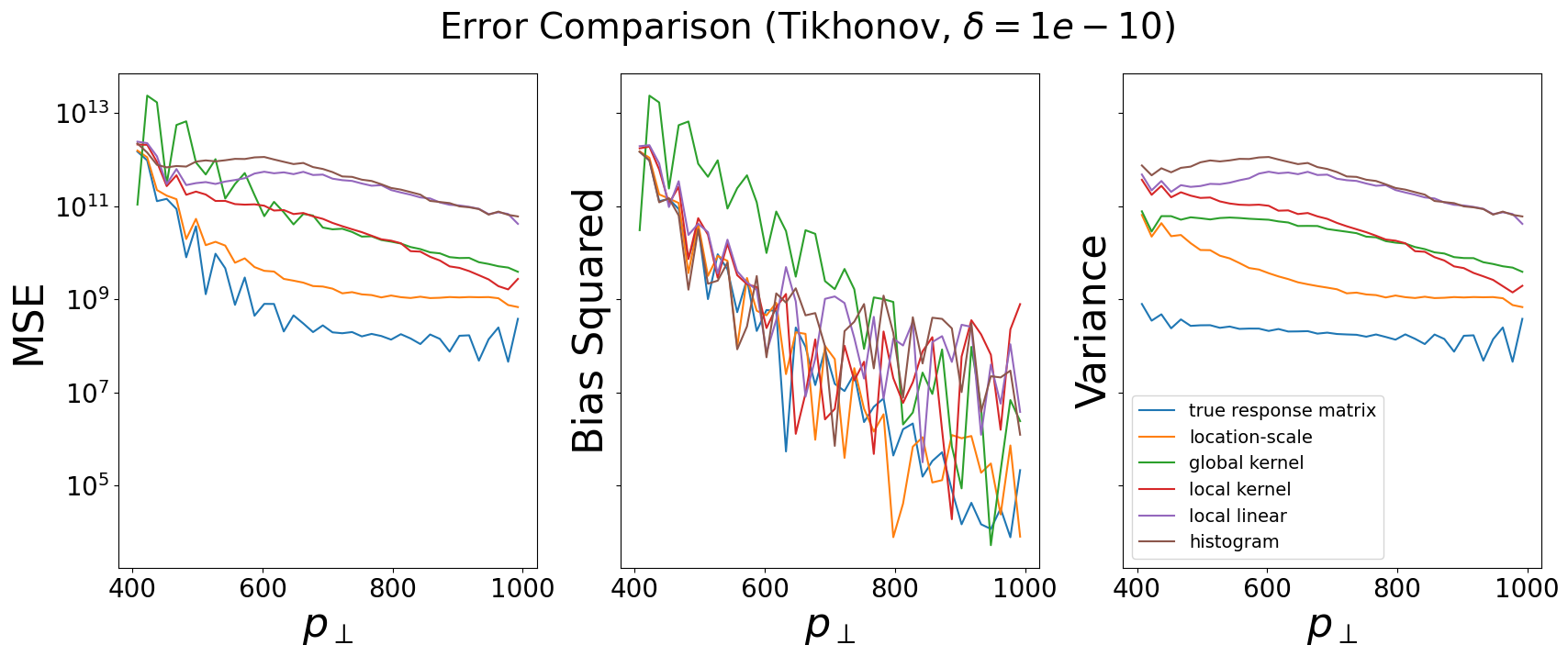}
        \caption{$\delta=10^{-10}$.}
        \label{fig:mse_tikhonov_delta=1e-10}
    \end{subfigure}

    \vspace{0.5em}

    \begin{subfigure}{\linewidth}
        \centering
        \includegraphics[width=\linewidth]{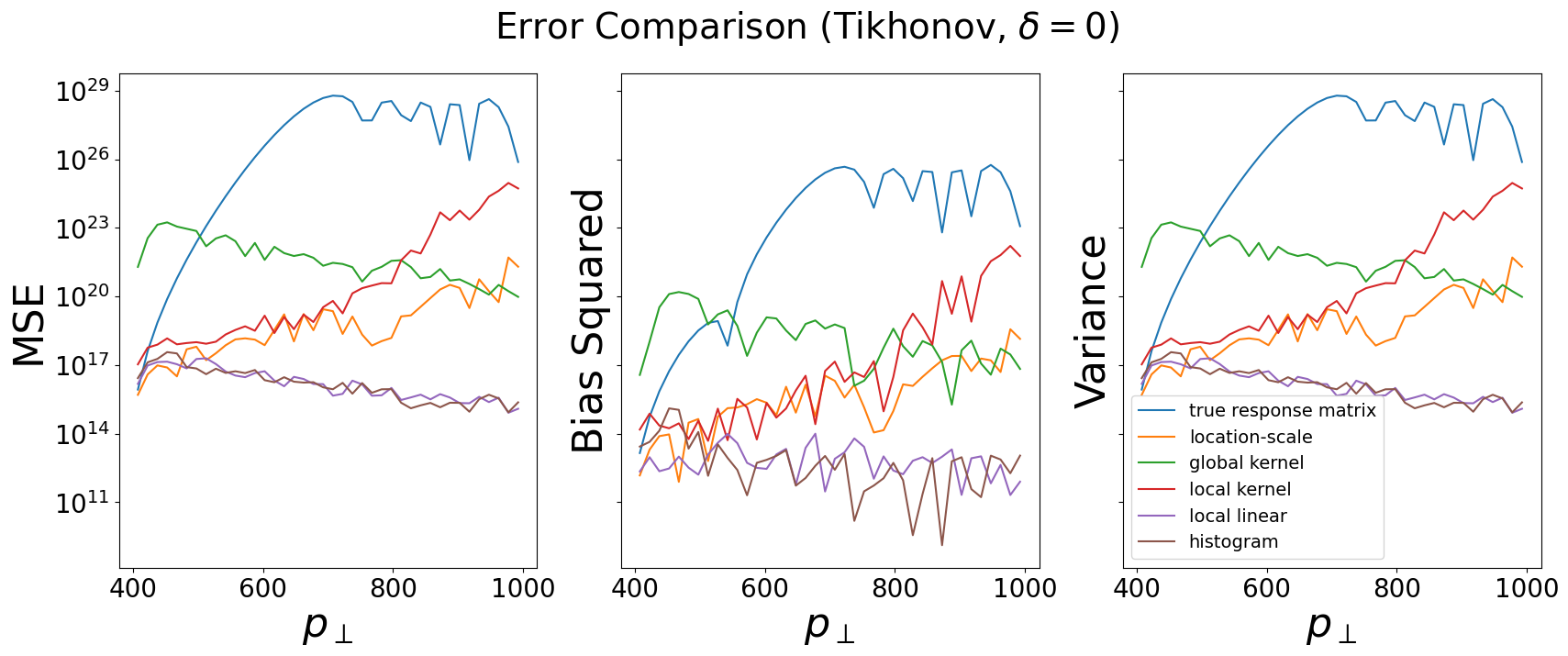}
        \caption{$\delta=0$ (unregularized solution).}
        \label{fig:mse_tikhonov_delta=0}
    \end{subfigure}

    \caption{Bin-wise mean squared error (left), squared bias (middle), and variance (right) for Tikhonov regularization with different $\delta$ values. All three panels for each $\delta$ share the same y-axis for direct comparison.}
    \label{fig:mse_tikhonov}
\end{figure}

An interesting observation is that when $\delta=0$ (Figures \ref{fig:tikhonov_sol_alpha=0} and \ref{fig:mse_tikhonov_delta=0}), i.e., without any regularization, the least-squares solution using the true response matrix performs surprisingly poorly. It has the largest MSE, bias and variance across most bins. Table \ref{table:condition_number} illustrates that the true response matrix has a notably larger condition number compared to the estimated response matrices. This explains the large variance of the solution with the true matrix. However, even with this extreme ill-conditioning, the least-squares solution should still, in principle, be unbiased. The non-negligible bias we nevertheless observe may be indicative of numerical artifacts arising from the ill-conditioned matrix inversion.

Moreover, the histogram estimate in this case is the best among all the methods, which can be explained by its smallest condition number. Even though it is a noisy and sub-optimal estimator of $\bm{K}$, it yields numerically the most stable solution for $\bm{\lambda}$. The reason for the small condition number is that a randomly perturbed matrix is likely to be well-conditioned.  Specifically, we can write the histogram estimate as the true response matrix plus some random perturbation, denoted as $\hat{\bm{K}}^{hist}=\bm{K}+\bm{N}$, where $\bm{N}$ is some random noise. \citet{Tao2007,Tao2008} showed that $\bm{K}+\bm{N}$ is well-conditioned with high probability even if $\bm{K}$ is ill-conditioned, which explains the reduced conditional number for the histogram estimator in Table~\ref{table:condition_number}.

As the regularization strength approaches 0, both bias and variance increase due to increased numerical artifacts and statistical variation. In such scenarios, the solution based on the true response matrix tends to get worse, while the solution by the histogram estimate starts to surpass the other methods due to the \emph{implicit regularization} provided by its moderate condition number. However, the overall quality of the unfolded point estimator still degrades significantly for all the methods as the regularization strength approaches zero. Therefore, in practice, using explicit regularization with $\delta > 0$ is still preferred, and in those cases, the other matrices beat the histogram method.

\begin{figure}[H]
    \centering
    \begin{subfigure}{\linewidth}
        \centering
        \includegraphics[width=\linewidth]{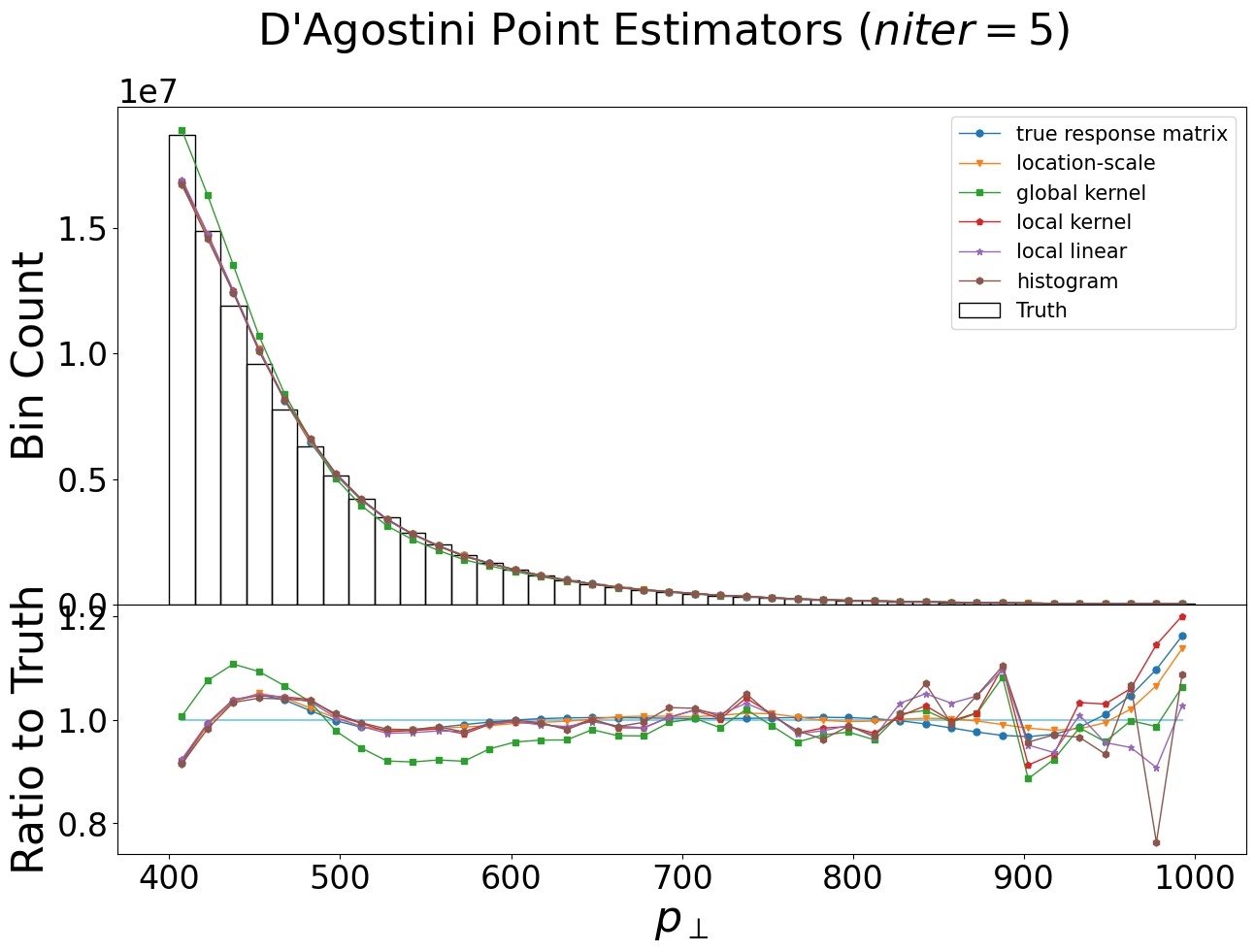}
        \caption{D'Agostini unfolding solution with $5$ iterations.}
        \label{fig:IBU_sol_niter=5}
    \end{subfigure}

    \vspace{0.8em}

    \begin{subfigure}{\linewidth}
        \centering
        \includegraphics[width=\linewidth]{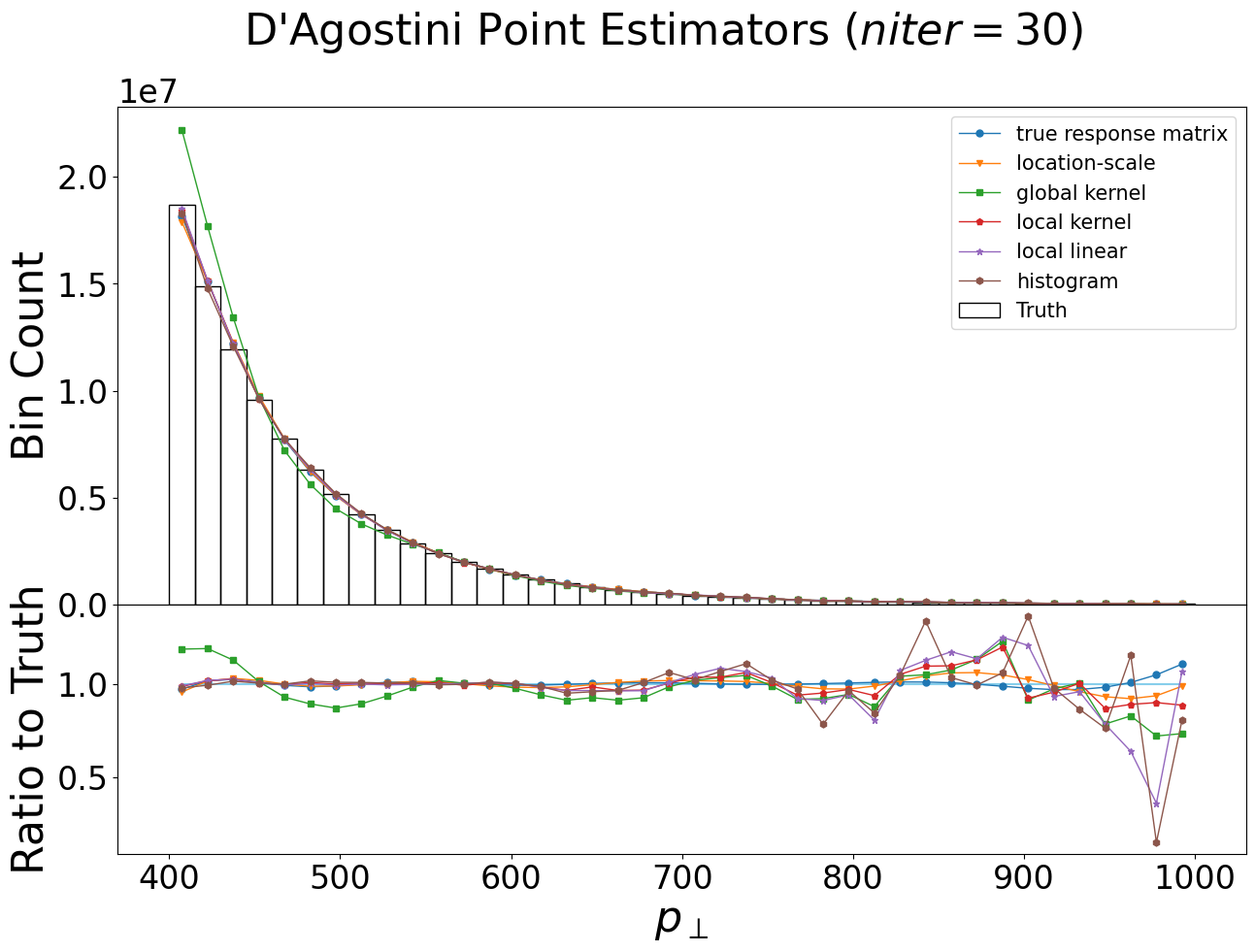}
        \caption{D'Agostini unfolding solution with $30$ iterations.}
        \label{fig:IBU_sol_niter=30}
    \end{subfigure}

    \caption{
    D'Agostini solutions with (a) $5$ iterations and (b) $30$ iterations.
    }
    \label{fig:IBU_solutions_5_30}
\end{figure}

\begin{figure}[h]
    \centering
    \includegraphics[width=\linewidth]{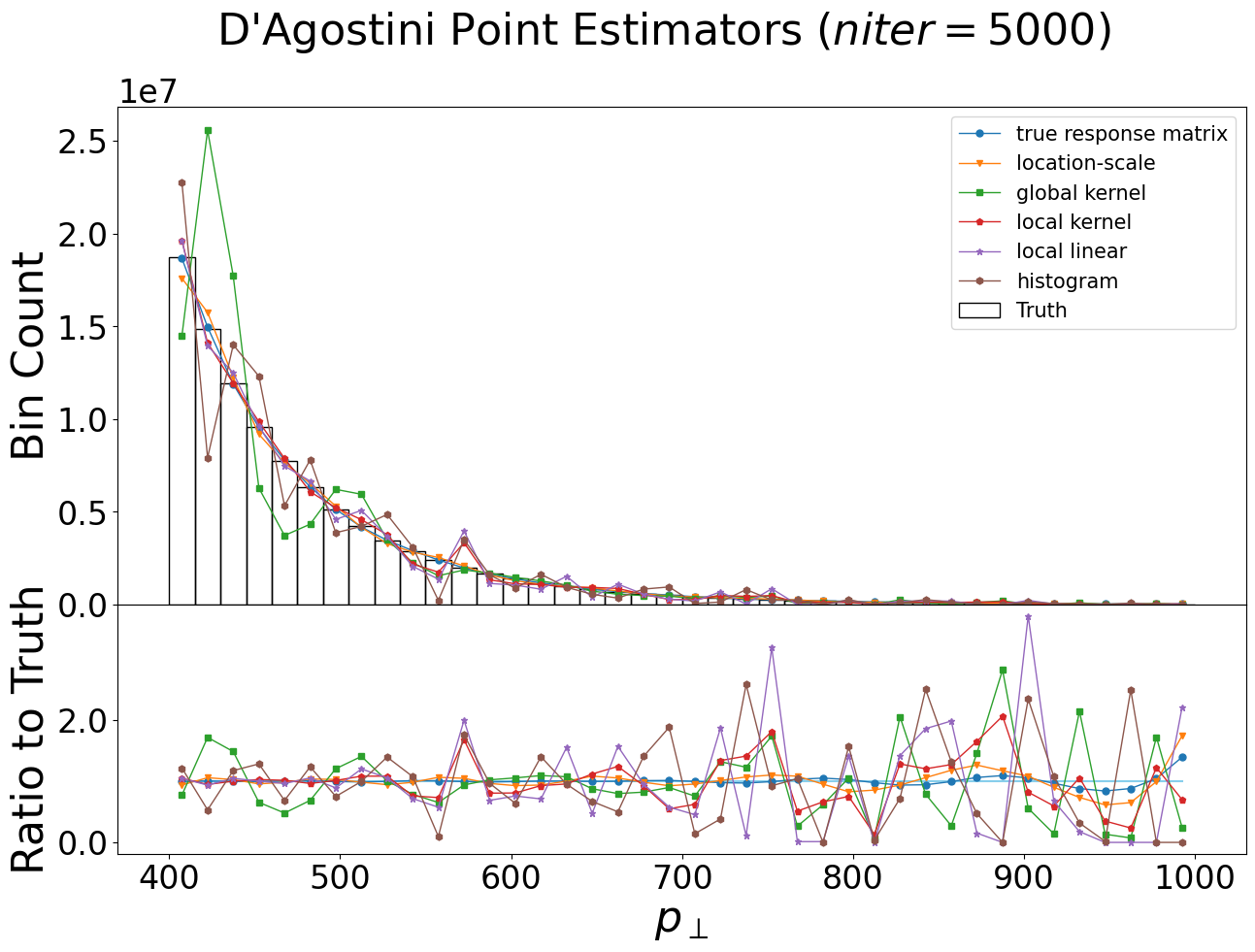}
    \caption{D'Agostini solution with $5{,}000$ iterations.}
    \label{fig:IBU_sol_niter=5000}
\end{figure}

\subsection{Numerical results for D'Agostini iteration}

\begin{figure}
    \centering
    \begin{subfigure}{\linewidth}
        \centering
        \includegraphics[width=\linewidth]{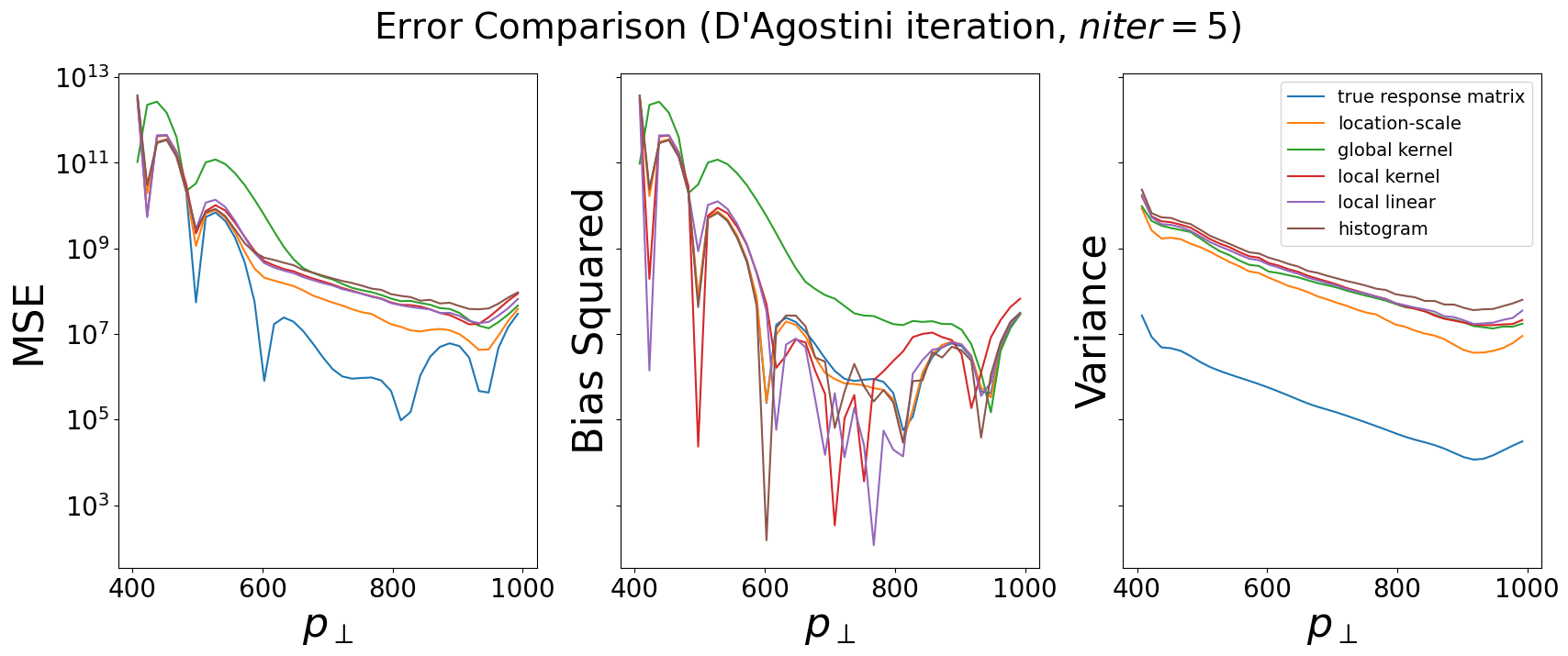}
        \caption{$niter=5$.}
        \label{fig:mse_IBU_sol_niter=5}
    \end{subfigure}

    \vspace{0.5em}

    \begin{subfigure}{\linewidth}
        \centering
        \includegraphics[width=\linewidth]{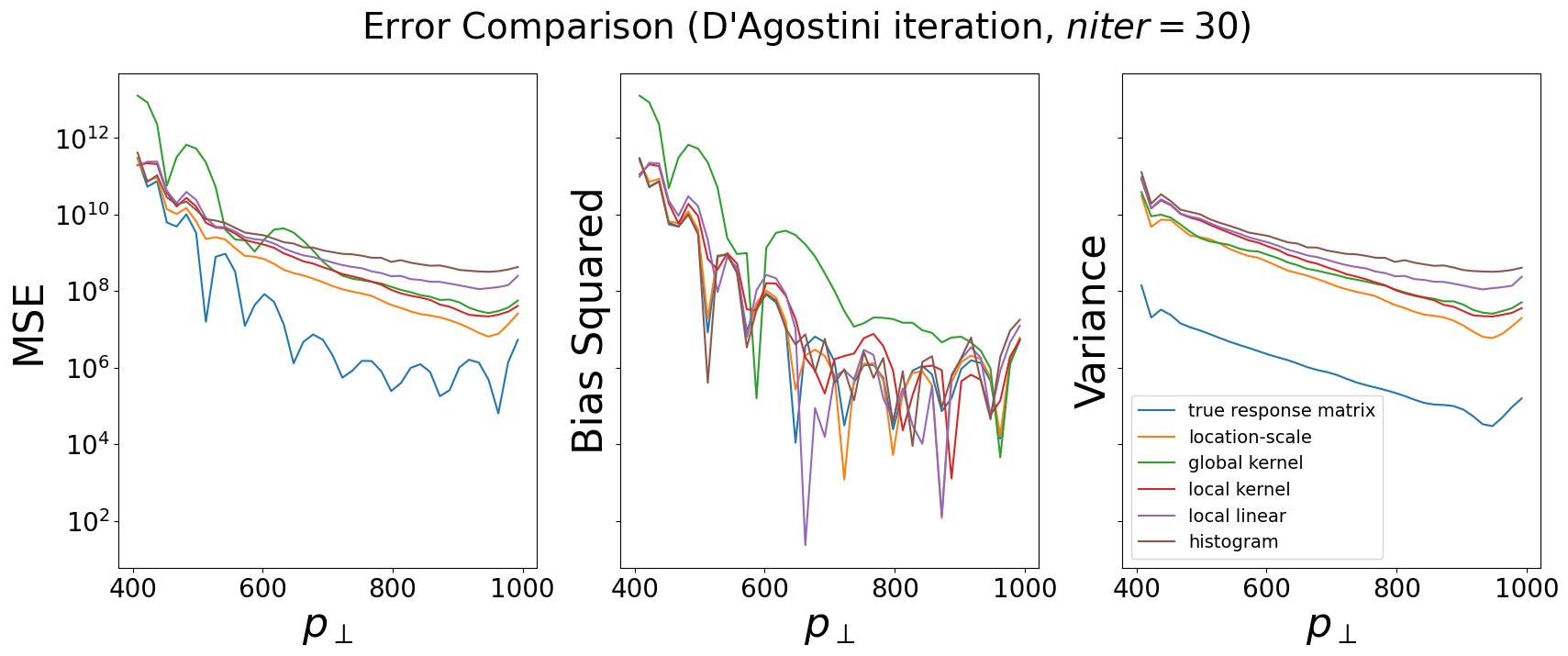}
        \caption{$niter=30$.}
        \label{fig:mse_IBU_sol_niter=30}
    \end{subfigure}

    \vspace{0.5em}

    \begin{subfigure}{\linewidth}
        \centering
        \includegraphics[width=\linewidth]{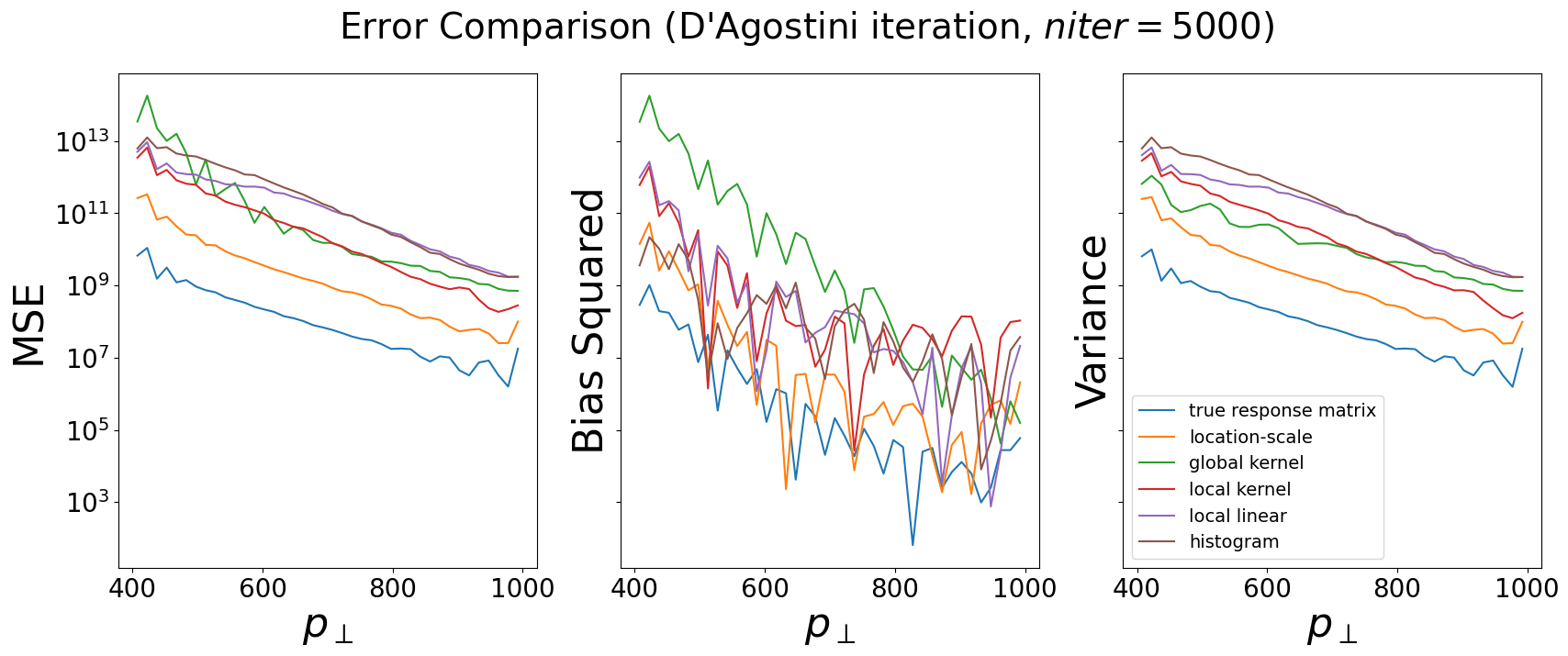}
        \caption{$niter=5000$.}
        \label{fig:mse_IBU_sol_niter=5000}
    \end{subfigure}

    \caption{Bin-wise mean squared error (left), squared bias (middle), and variance (right) for D'Agostini solutions with different iteration counts. All three panels for each $niter$ share the same y-axis for direct comparison.}
    \label{fig:mse_IBU}
\end{figure}

Figures \ref{fig:IBU_solutions_5_30}--\ref{fig:IBU_sol_niter=5000} illustrate one D'Agostini solution with various numbers of iterations for the different response matrix estimators. Figure \ref{fig:mse_IBU} shows the corresponding (log-scale) MSE, squared bias and variance for $M=1{,}000$ realizations. As in the case of Tikhonov regularization, the MSE of the D'Agostini solutions is also mostly driven by the variance. The relative magnitude of the variance aligns with the performance of each response matrix estimation method, with a better estimated response matrix generally leading to lower variance. But the difference in the variances is not as significant as in the Tikhonov case when the number of iterations is modest (Figures \ref{fig:mse_IBU_sol_niter=5} and \ref{fig:mse_IBU_sol_niter=30}). This indicates that the statistical uncertainty from the estimated response matrix is overshadowed by the strong regularization in this case.
However, as the number of iterations increases (Figure \ref{fig:mse_IBU_sol_niter=5000}), we see that the difference in the variances becomes more pronounced.

In both the Tikhonov and D'Agostini solutions, the bias of the global kernel method is significantly higher than for the other methods in the low $p_\perp$ region. This is because the estimated response matrix has a large bias in this region, as observed in Section~\ref{sec:compare_matrix}. As we discussed above, there does not seem to exist a single pair of global bandwidths that can produce a good estimate for all $p_\perp$. In this case, the chosen bandwidths oversmooth in the low $p_\perp$ region, resulting in a high bias for the estimated matrix which is then also reflected in the unfolded point estimate. Similarly, for the local linear method, the variance is larger than for the other methods (except for the naive histogram method) in the high $p_\perp$ region. This is due to the fact that the chosen bandwidths undersmooth in the high $p_\perp$ region. On the other hand, by using local bandwidths, the local kernel method provides a ``best-of-both-worlds'' approach by having, in most cases, bias comparable to the local linear method and variance comparable to the global kernel method.

\section{Application to Drell-Yan + jets events}
\label{sec:Drell-Yan}

\begin{figure*}[t]
    \centering
    \includegraphics[width=\textwidth]{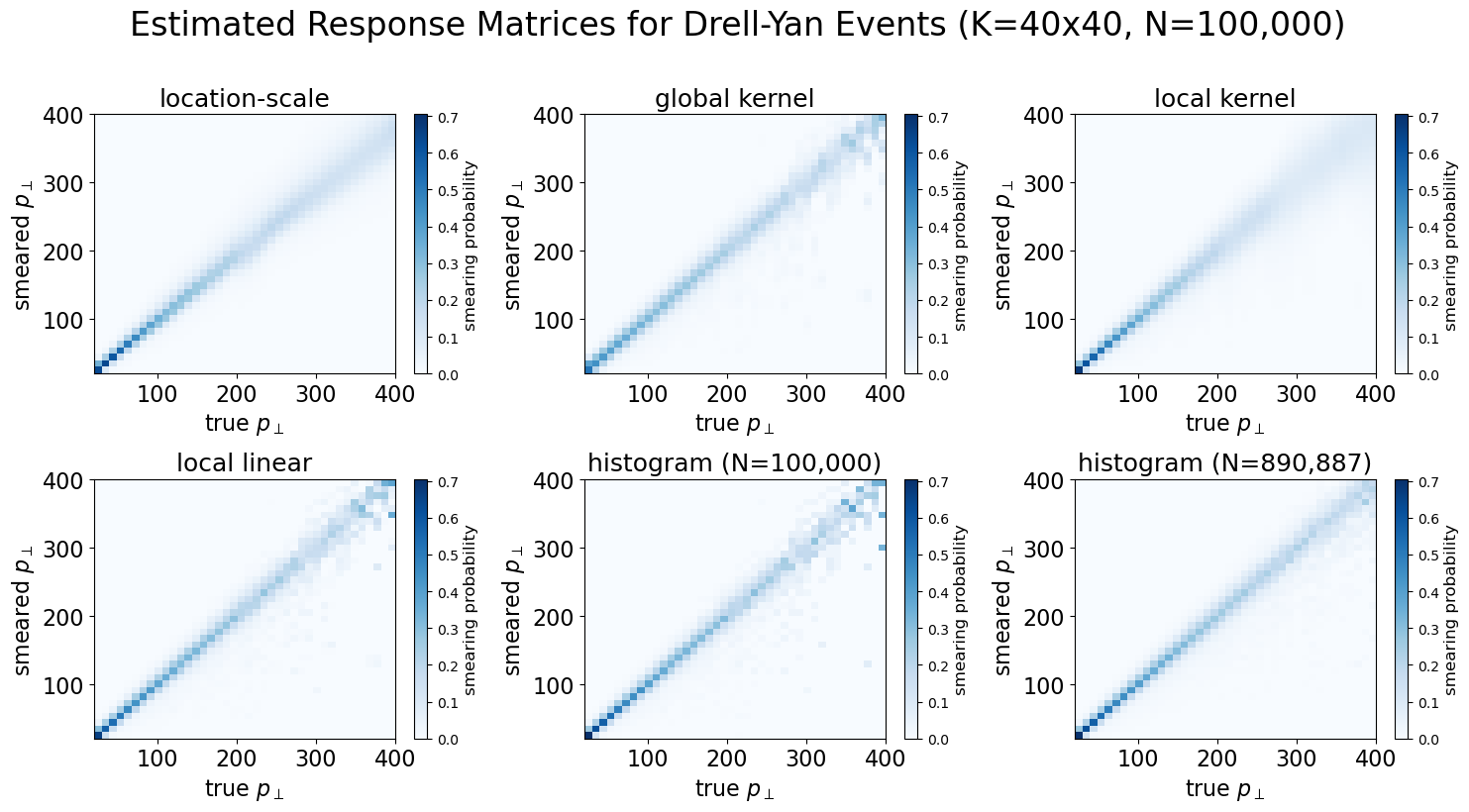}
    \caption{Estimated $40 \times 40$ response matrices using the different methods with one Monte Carlo simulation. The sample size of the Monte Carlo simulation is 100{,}000, except for the histogram method in the lower right panel, which uses the full sample size of 890{,}887.}
    \label{fig:Drell-Yan-mat_K=40x40}
\end{figure*}

We apply the methods to unfold the jet transverse momentum spectrum in simulated Drell-Yan events with jets produced in $pp$ collisions at center of mass energy of $\sqrt{s} = 13$ TeV. 
The matrix element is generated with MadGraph5 aMC@NLO (v2.9.9) at leading order in perturbative quantum chromodynamics (QCD), with up to two partons in the final state, and merged using the MLM scheme \cite{Alwall2008}. The parton density functions are taken from NNPDF 3.0 \cite{NNPDF2017}. Parton showering and hadronization are simulated with Pythia 8.2 \cite{Alwall2008,Alwall2014,Sjostrand2015}. The detector response is modeled with Delphes using a description of the CMS detector \cite{Delphes2014,CMS2008}. For both the particle-level and detector-level data, we include events that contain two muons and di-muon invariant mass within $91\pm 20 \text{ GeV}$. Jets are required to have $|\eta|\leq 2.5$, and for each muon, the angular separation $\Delta R$ between the muon momentum vector and each jet momentum vector must exceed 0.4. The remaining events have sample size of 890{,}887 which we split into a training set with 100{,}000 events and a test set with 790{,}887 events. Both training and test sets have pairs of particle-level and detector-level data. The training set represents the Monte Carlo data that is used to estimate the response matrix. The test set represents the experimental detector-level data that is used for unfolding, while the corresponding particle-level data, which is not observable in practice, is used for validation.  We consider the range of both the particle- and detector-level histograms to be $[T_{\min},T_{\max}]=[S_{\min},S_{\max}]=[20 \text{ GeV},400 \text{ GeV}]$ and use 40 equal-sized bins in both histograms.

Figure \ref{fig:Drell-Yan-mat_K=40x40} shows the estimated response matrices obtained using different conditional density estimation methods. The response matrices are estimated using a sample size of $N=100{,}000$, except for the histogram method shown in the lower right panel, which uses the full sample of 890{,}887 events. With the same sample size of $N=100{,}000$, the histogram estimate is noticeably noisier compared to the CDE methods, particularly in the sparsely sampled right tail of the spectrum. There are also differences between the various CDE methods that broadly align with their respective behavior in the inclusive jet study in Section~\ref{sec:compare_matrix}. However, since the true response matrix is unknown here, it is not possible to directly assess their accuracy, but the full-sample histogram estimate may serve as a proxy for the true response matrix, given that it is an unbiased estimator and benefits from reduced statistical fluctuations. If we take the full-sample histogram estimate as the ground truth, then the local linear method appears to provide the most accurate estimate of the response matrix. In general, all the estimated response matrices show that the smearing probabilities are more sharply concentrated near the diagonal for low $p_\perp$, reaching values as high as 0.7, while in the high $p_\perp$ region, the smearing becomes more dispersed. Overall, the degree of smearing does not seem to be as severe here as in the previous inclusive jet case study.

\begin{figure}[h]
    \centering
    \begin{subfigure}{\linewidth}
        \centering
        \includegraphics[width=\linewidth]{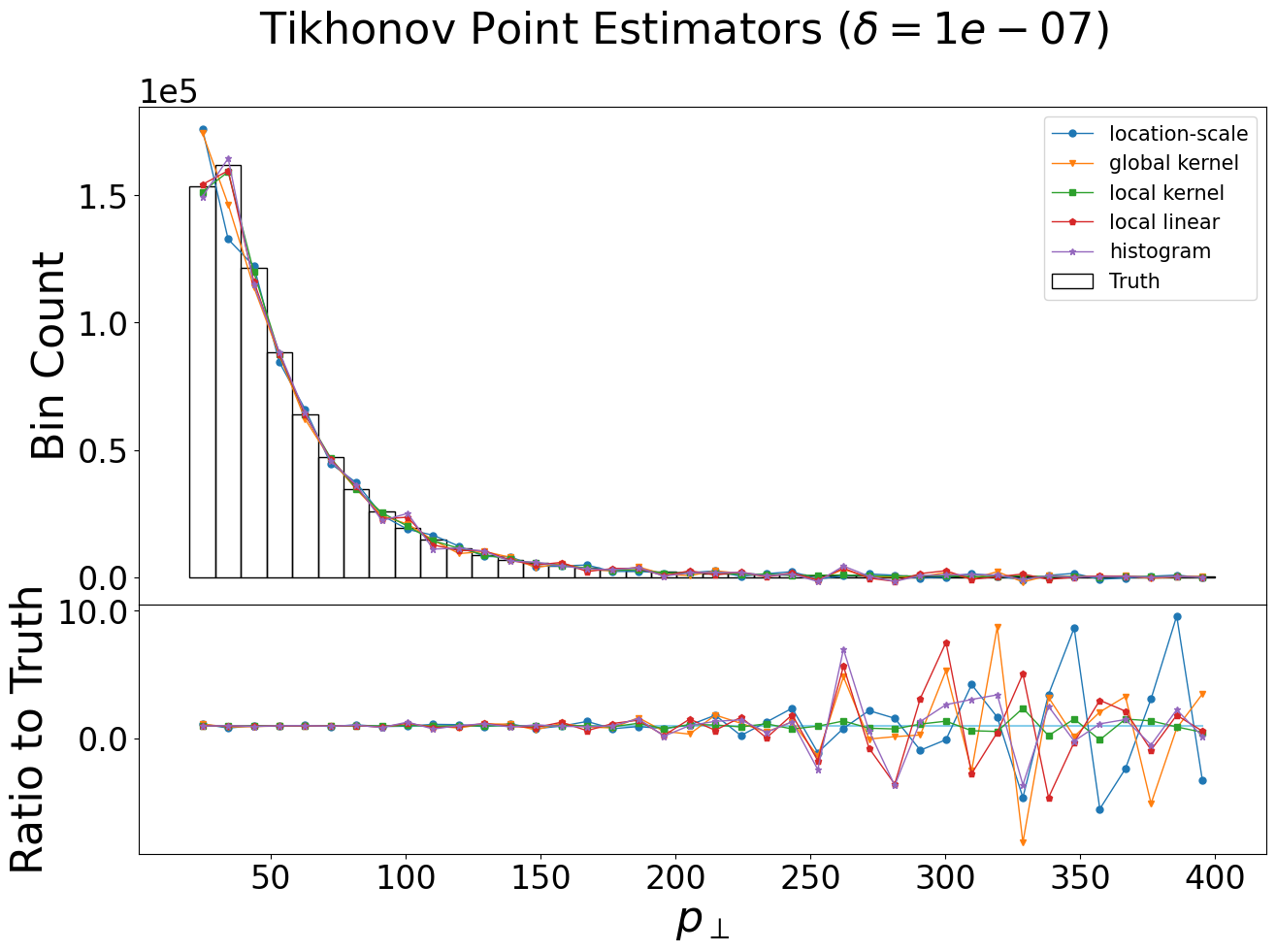}
        \caption{Tikhonov-regularized solution with $\delta=10^{-7}$.}
        \label{fig:Drell_Yan_least_squares_sol_alpha=1e-7}
    \end{subfigure}

    \vspace{0.8em}

    \begin{subfigure}{\linewidth}
        \centering
        \includegraphics[width=\linewidth]{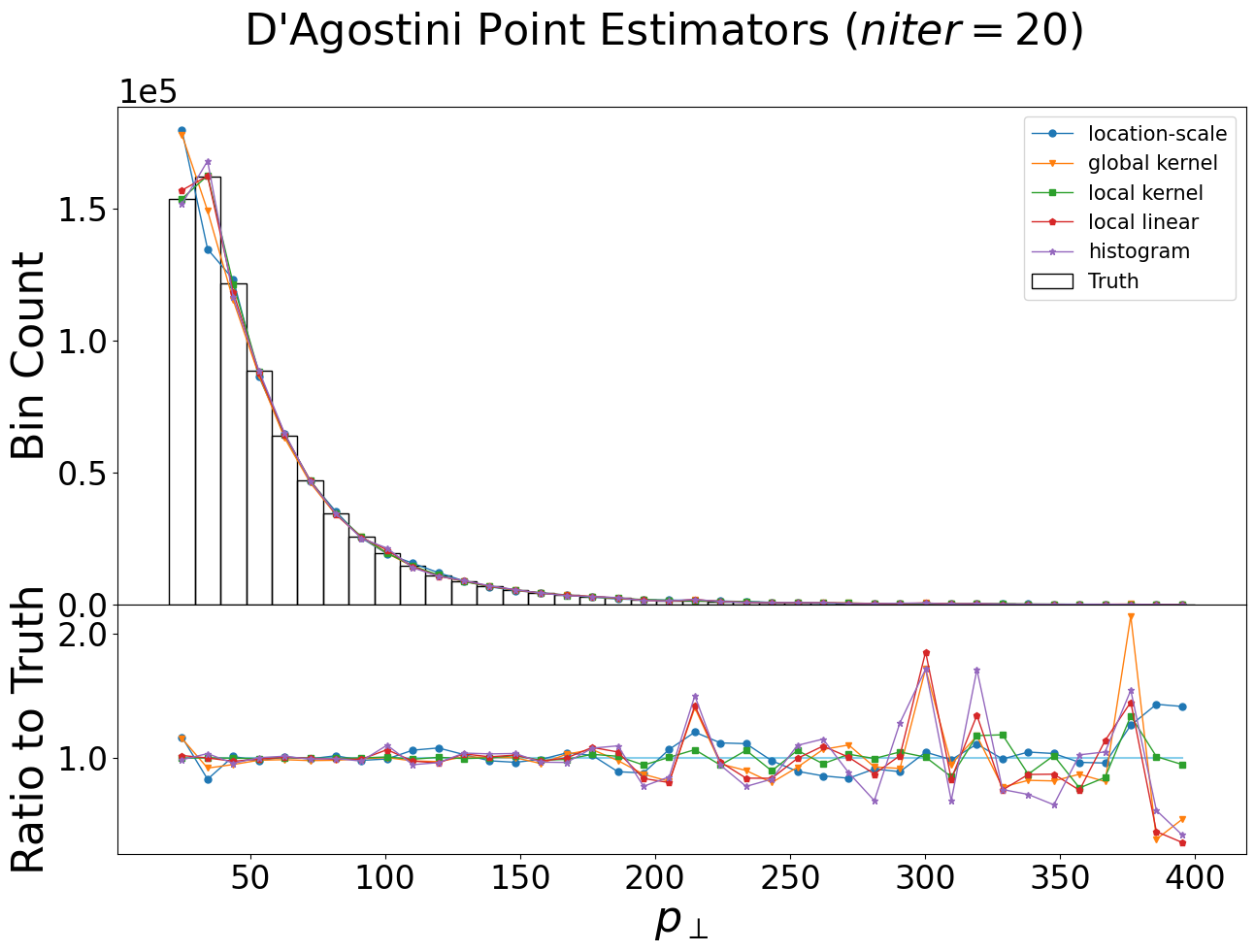}
        \caption{D'Agostini solution with 20 iterations.}
        \label{fig:Drell_Yan_IBU_sol_niter=20}
    \end{subfigure}

    \caption{Comparison of unfolding methods on Drell-Yan data: (a) Tikhonov regularization and (b) D'Agostini iteration.}
    \label{fig:Drell_Yan_unfolding_comparison}
\end{figure}

Figures \ref{fig:Drell_Yan_least_squares_sol_alpha=1e-7} and \ref{fig:Drell_Yan_IBU_sol_niter=20} show the Tikhonov-regularized solution with $\delta=10^{-7}$ and D'Agostini solution with 20 iterations, respectively. The regularization parameters ($\delta$ and $niter$) are selected empirically to best balance the bias-variance tradeoff. Since we do not have access to the ground truth, the proxy truth histogram in this case is the empirical particle-level histogram, which itself has a statistical uncertainty. However, the sample size is chosen to be large to ensure that this uncertainty remains small. The results show some similar characteristics to those discussed in Section \ref{sec:effect_on_unfold_spectrum}. In particular, when using the D'Agostini solution with a moderate number of iterations, the differences between the different response matrix estimators are generally less pronounced than in the Tikhonov case. In contrast, the Tikhonov-regularized solution shows greater variance, particularly in the high $p_\perp$ region, as evidenced by the larger fluctuations in the ratio plot shown in Figure \ref{fig:Drell_Yan_unfolding_comparison}.

Interestingly, both the location-scale and global kernel estimators deviate substantially from the other methods in the low $p_\perp$ region under both solutions. For the global kernel method, this behavior aligns with what was observed in the inclusive jet study, where a large bias in the estimated response matrix in the low $p_\perp$ region led to poor performance. However, the location-scale method, which performed well in the inclusive jet study, underperforms in the present setting. This discrepancy might be caused by the location-scale assumption being violated in this more realistic setting.

On the other hand, the local kernel method with adaptive local bandwidths mitigates the issue in the low $p_\perp$ region, resulting in a more stable performance across the spectrum. The local linear method also yields similar results, though with more fluctuation in the high $p_\perp$ region in the Tikhonov solution, likely due to undersmoothing as discussed in Section \ref{sec:effect_on_unfold_spectrum}. The histogram method, while not as smooth as the local kernel method, still provides a comparable estimate for the unfolded solution --- possibly due to the relatively mild smearing observed in the estimated response matrices and the implicit regularization phenomenon discussed in Section~\ref{sec:implicit_reg}. Nonetheless, it is important to note that the results shown represent a single realization and a more comprehensive evaluation would require access to multiple repetitions of this experiment.

\section{Discussion and conclusion}
\label{sec:discussion}
In this work, we proposed a new class of methods for estimating the response matrix in particle unfolding. 
Rather than the traditional histogram method based on directly binning the events, we first estimate the response kernel on the unbinned space. Then, an estimator for the response matrix can be obtained by plugging the estimated response kernel back into the definition of the response matrix. One benefit of this approach is that it provides a smoother, statistically more efficient estimator for the response matrix if the true response kernel is relatively smooth, which mitigates the issue of statistical noise in the histogram approach.

Estimating the response kernel is equivalent to the well-studied statistical problem of conditional density estimation and we considered and compared several non-parametric CDE methods in this task. The resulting response matrices and unfolded solutions based on the CDE approach show better, or at least no worse, performance than the histogram approach in most cases. Moreover, a better estimate for the response kernel/matrix generally leads to a better unfolded solution, which aligns well with our intuition. An intriguing exception is the case of unregularized unfolding in which we uncovered an unexpected implicit regularization phenomenon caused by the noise in the histogram estimate.

There are limitations to be improved in future work. First, bandwidth selection is inevitable in nonparametric conditional density estimation. The normal reference rule used here is computationally efficient, but it may be suboptimal when the normality assumption is violated---although it can still yield reasonable results in non-normal settings.
More sophisticated data-driven methods based on cross-validation or bootstrap techniques have been proposed \citep{Fan2004, Ruppert1997}, but they tend to be computationally expensive, and in our preliminary experiments, we did not observe substantial improvements by using such methods. An important avenue for future work is the uncertainty quantification of the unfolded solution in the presence of statistical uncertainty in the response matrix. Thus far, we have only considered point estimates based on the estimated response matrix. However, it is not entirely clear how to best incorporate the uncertainty of the response matrix into the uncertainty of the unfolded solution. Standard software \cite{RooUnfold} handles this through bootstrapping or nuisance parameters, but to the best of our knowledge, the statistical properties of these uncertainties have not been systematically studied and it is not immediately clear if these methods apply to the CDE estimators where the smoothing makes the response matrix bins correlated. Lastly, many recent works have proposed to perform unfolding without binning using machine learning methods \citep{Andreassen2020, Diefenbacher2024, Backes2024, Shmakov2023, Datta2019, Bellagente2020}. Interestingly, these methods bypass the problem of explicitly estimating the response kernel/matrix. Instead, they directly produce the unfolded solution using only the MC training pairs $\{(X_i,Y_i)\}_{i=1}^n$ and the smeared observations without requiring a plug-in estimate of the response kernel. A rigorous understanding of the statistical properties and tradeoffs of these methods is an important topic for future work.


\section*{Acknowledgments}
This work was supported by NSF grants PHY-2020295, DMS-2053804 and DMS-2310632.





\bibliographystyle{elsarticle-num-names} 
\bibliography{biblio}





\end{document}